\documentclass{aa}  

\usepackage{graphicx}
\usepackage{txfonts}

\usepackage{hyperref}
    \hypersetup{colorlinks=true,linkcolor=blue,citecolor=blue,urlcolor=blue}
    \makeatletter
    \renewcommand*\aa@pageof{, page \thepage{} of \pageref*{LastPage}}
    \makeatother
\usepackage{algorithm} 
\usepackage{algpseudocode}
\usepackage{gensymb}

\begin{document}

   \title{SOAP-GPU: Efficient Spectral Modelling of Stellar Activity Using Graphical Processing Units}
   
   \titlerunning{SOAP-GPU}
   \authorrunning{Yinan Zhao et al.}

   \author{Yinan Zhao
          \inst{1}
          \and
          Xavier Dumusque\inst{1}
          }

   \institute{Department of Astronomy of the University of Geneva, 51 chemin de Pegasi, 1290 Versoix, Switzerland\\
              \email{yinan.zhao@unige.ch} }

 
 
  \abstract
   {Stellar activity mitigation is one of the major challenges for the detection of earth-like exoplanets in radial velocity measurements. Several promising techniques are now investigating the use of spectral time-series, to differentiate between stellar and planetary perturbations. In this context, developing a software that can efficiently explore the parameter space of stellar activity at the spectral level is of great importance.}
   {The goal of this paper is to present a new version of the Spot Oscillation And Planet (SOAP) 2.0 code that can model stellar activity at the spectral level using graphical processing units (GPUs).}
   {We take advantage of the computational power of GPUs to optimise the computationally expensive algorithms behind the original SOAP 2.0 code.
     For that purpose, we developed GPU kernels that allow to model stellar activity on any given wavelength range. 
   In addition to the treatment of stellar activity at the spectral level, SOAP-GPU also includes the change of spectral line bisectors from center to limb, and can take as input PHOENIX spectra to model the quiet photosphere, spots and faculae, which allow to simulate stellar activity for a wider space in stellar properties. }
   {Benchmark calculations show that for the same accuracy, this new code improves the computational speed by a factor of 60 compared with a modified version of SOAP 2.0 that generates spectra, when modeling stellar activity on the full visible spectral range with a resolution of R=115'000. 
   Although the code now includes the variation of spectral line bisector with center-to-limb angle, the effect on the derived RVs is small. We also show that it is not possible to fully separate the flux from the convective blueshift effect  when modeling spots, due to their lower temperature and thus the appearance of molecular absorption in their spectra. Rather negligible for the Sun, this degeneracy between the flux and convective blueshift effect become more important when we move to cooler stars, however, this issue does not impact the estimation of the total effect (flux plus convection), and therefore users can trust this output.}
 {The publicly available SOAP-GPU code allows to efficiently model stellar activity at the spectral level, which is essential to test further stellar activity mitigation techniques working at the level of spectral timeseries not affected by other sources of noise. Besides a huge gain in performance, SOAP-GPU also includes more physics and is able to model different stars than the Sun, from F to K dwarfs, thanks to the use of the PHOENIX spectral library. We however note that due to the limited understanding of stellar convection and activity on other stars than the Sun, the more we go away from the solar case, the more the output of the code should be taken with care.}

   \maketitle
%

\section{Introduction}
The radial velocity (RV) method has been proven to be one of the most successful method to detect exoplanets since the discovery of the first exoplanet orbiting a solar-type star \citep{Mayor-1995Nature}. In order to detect earth-like planets orbiting in the habitable zone of its parent star, a precision of a few dozens of $\rm{cm s^{-1}}$ must be reached. Although the state-of-the-art spectrographs such as ESPRESSO, and EXPRESS are not far from that precision \citep[50 and 58 $\rm{cm s^{-1}}$, respectively][]{Pepe-2021,Brewer:2020aj}, the main limitation to detect Earth-like planets with the RV technique is stellar activity. Two major physical processes dominating stellar activity on a time scale of the host star's rotational period are the flux imbalance due to the temperature difference and therefore contrast between active and quiet regions (hereafter flux effect. \citep[e.g.][]{Saar-1997ApJ, Dumusque-2014b, Donati-2017MNRAS}) and the inhibition of convective blueshift (hereafter CB effect). The CB effect is due to the presence of strong local magnetic fields inside active regions, which suppress the CB inside those regions and leads to positive RV variations \citep[e.g.][]{Cavallini-1985,Meunier-2010a}. 

Many methods have been proposed to mitigate activity-induced variations using photometric and spectroscopic time series. In the one-dimensional time series space, many parametric models based on analytic forms or different Gaussian process (GP) frameworks have been developed to model stellar activity using photometry or spectroscopic activity indicators \citep[e.g.][]{Aigrain-2012, Rajpaul-2015MNRAS, Aigrain-2016MNRAS, Gilbertson-2020ApJ, Barragan-2022MNRAS}.
Jointly modeling the data with Keplerians to model planets in addition to a GP to model stellar activity may significantly reduce the stellar activity but may also lead to overfitting when the GP kernel or priors are not wisely set. This is particularly dangerous when the planetary properties are not constrained from transit observations.

Due to inherent problems in modeling stellar activity in one-dimensional time series, the community is now shifting toward modeling it in a two-dimensional space. 
\cite{Collier-Cameron-2021MNRAS} calculated the autocorrelation function (ACF) of cross-correlation function \citep[hereafter CCF][]{Baranne-1996}, to isolate Doppler shift from shape shift variations and applied principle component analysis (PCA) on the obtained ACFs to model shape changes related to stellar activity. A planet signal of amplitude $\sim40\,\rm{cm/s}$ can be recovered when the algorithm is applied to the HARPS-N solar data \citep[][]{Dumusque-2015ApJ,Collier-Cameron-2019MNRAS,Dumusque-2021aa}. \cite{Zhao-2022arXiv} projected CCFs time series onto the Fourier basis functions and modelled line variability using different basis. Results on simulated data show a 48$\%$ reduction in RV rms. \cite{Beurs-2020arXiv} trained a convolutional neural network (CNN) on both simulated CCFs and HARPS-N solar CCFs and were able to significantly reduce stellar activity effects.
 
The idea behind building the CCF is to extract with the best precision the RV information contained in a spectrum. However, key variations at the spectral level related to stellar activity may be lost when performing the dimensionality reduction imposed by the CCF. Therefore, several methods have been proposed to disentangle stellar activities at the spectral level. \cite{Davis-2017ApJ} applied PCA to simulated spectral time series and demonstrated that eigen-vectors are spectral line dependent. \cite{Rajpaul-2020MNRAS} used GP to directly derive RV information from spectral time series. \cite{Jones-2017arXiv} also applied multivariate GP to model stellar activity on PCA-reduced spectral dataset. \cite{Cretignier-2022aa}, based on the knowledge that the impact of stellar activity is line-depth dependant \citep[][]{Cretignier:2020aa}, used PCA to model stellar activity in the flux-flux gradient space (named the ``shell'' space) and results on HD10700 ($\tau$ Ceti) and HD12861 ($\alpha$ Cen B) indicates the method can successfully remove variations from non-Doppler origin. Last by not least, \cite{Binnenfeld-2020aa,Binnenfeld-2022aa} are developing the unit-sphere representation periodogram (USuRPER), to seperate Doppler from other RV variations. This technique is based on representing spectra as unit vectors in a multidimensional hyperspace.

The spectral time series used to evaluate the performance of the algorithms developed to mitigate stellar activity at the spectral level are either obtained from simulated data or real observations. 
The major issue with simulations, is that most of them only model the RV activity effect at the CCF level \citep[e.g.][]{Dumusque-2014b,Herrero:2016aa} due to computational inefficiency. A few other libraries of simulated spectra affected by stellar activity exist, but generating them takes hours to run, which is not convenient when exploring the parameter space in stellar activity and properties \citep[e.g.][]{Gilbertson:2020aa, Dumusque-2016aa}.
Regarding real observations, solar data obtained by the HARPS-N solar telescope \citep[][]{Collier-Cameron-2019MNRAS,Dumusque-2021aa}, HELIOS on HARPS\footnote{\url{https://www.eso.org/public/announcements/ann18033/}} and more recently the solar feed of NEID \citep[][]{Lin:2021aa} are the best we can get, in terms of S/N and sampling. However, those spectra corresponds for the most part to quiet activity phases of the Sun (end of cycle 24 end beginning of cycle 25) and can only used to mitigate stellar activity for star very similar to the Sun. When moving to stellar observations, the recent Extreme precision Spectrograph (EXPRES) Stellar Signals Project (ESSP) shared some valuable data. However, due to the small number of stars and the rather small number of spectra available, it was rather difficult to compare different activity mitigation techniques together \citep{Zhao-2022AJ}. As a conclusion of this discussion, it is essential for the community to have access to a code that can simulate efficiently stellar activity at the spectral level, and for a wide range of stellar properties.

In this paper, we present a new code, Spot Oscillation And Planet Graphical Process Unit (SOAP-GPU) based on GPU computation that can efficiently model simplified and realistic stellar activity at the spectral level. In Sect \ref{sec2}, we revisit the architecture of the SOAP 2.0 code it is based on \citep{Dumusque-2014b} and discuss about its limitations. The algorithms behind SOAP-GPU are presented in Sect \ref{sec3}. In Sect \ref{sec4}, we explore the physical parameters of stellar activity and simulation of different cases are presented. Finally, we draw our conclusion in Sect \ref{sec5}. The SOAP-GPU code is publicly available on Github and Zenodo\footnote{code available here \url{https://github.com/YinanZhao21/SOAP_GPU} and \url{https://doi.org/10.5281/zenodo.7499461}} along with a brief manual and some examples.

\section{Revisiting SOAP 2.0}\label{sec2}
    In this section, we revisit the code Spot Oscillation And Planet \citep[SOAP 2.0][]{Dumusque-2014b}. This code aims at modeling both the flux effect and the CB effect of active regions affecting RV measurements. Although the public version of the SOAP 2.0 code can only simulate stellar activity at the level of the CCFs, modeling the effect at the spectral level follow the same ideas. In this section, we first discuss the basic algorithms behind SOAP 2.0 and demonstrate the limit of the code, in terms of computational efficiency, when we want to model stellar activity at the spectral level.
  
    \begin{figure}[htbp]
   \includegraphics[scale=0.25]{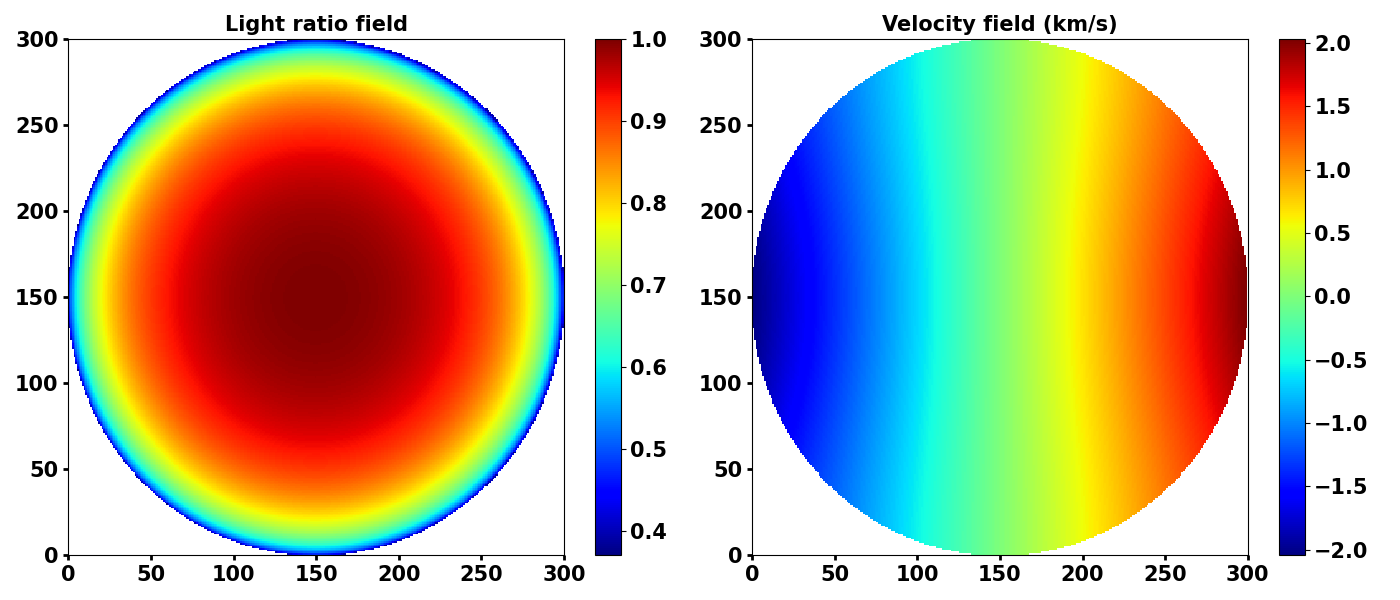}
   \caption{The stellar disk is initialized with velocity
   and intensity fields. \emph{Left:} The intensity in each cell is computed depending on a limb darkening law. \emph{Right:} The velocity in each cell is computed considering rotational period, stellar inclination and radius of the star. As we can see, iso-velocity lines are not vertical as we implemented differential rotation in SOAP-GPU, which was not the case in SOAP 2.0.
   }
   \label{Fig0}%
   \end{figure}  
\subsection{The structure of SOAP 2.0}
SOAP 2.0 first computes the ``quiet'' (without any active region) emission spectrum of the star. To do so, a 2-dimension stellar disk containing N $\times$ N cells is initialized \citep[N being the resolution of the disk, the same parameter called ``grid'' in][]{Boisse-2012b}. Velocity and intensity of all disk cells are computed based on the physical configuration of the star (rotational period, stellar inclination and radius of the star) and a limb darkening law (as shown in Fig.~\ref{Fig0}). In each cell, the quiet photosphere spectrum is injected, weighted by the cell intensity (limb-darkening), and Doppler-shifted to the projected velocity of that cell (rotation). Linear interpolation is applied at this step to project the Doppler-shifted spectrum into the original wavelength grid, to make sure that spectra in different cells are on a common wavelength grid. We note that the public version of SOAP 2.0 was using the CCF of the high-resolution Kitt Peak Observatory Fourier Transform Spectrograph (FTS) quiet photosphere spectrum \citep[$S_{quiet}(\lambda)$, ][]{Wallace-1998} as approximation of the quiet Sun to increase computational speed. However, injecting the original spectrum is possible, with the only difference that the dimension of the input is $\sim$500000, compared to 400 for the CCF, and that we will need to apply a Doppler-shift each time we want to change the velocity of this spectrum, while a simple translation was sufficient in the case of the CCF. After injecting the quiet photosphere spectrum in each cells, the integrated quiet solar spectrum is obtained by summing the content of all the cells together. All those processes are summarized in the pseudo code below (Algorithm 1).
\begin{algorithm}
	\caption{Quiet spectrum integration} 
	\begin{algorithmic}[1]
		\For {$X_{location}=1,2,\ldots N$}
			\For {$Y_{location}=1,2,\ldots,N$}
			    \State Shift $S_{quiet}(\lambda)$ with velocity $\mathbf{vel}_{X,Y}$. 
			    $S_{quiet}(\lambda)\rightarrow\ S_{quiet}(\lambda^{'})$.
			    \State Do linear interpolation to project the spectrum back to the original wavelength grid.
			    $S_{quiet}(\lambda^{'})\rightarrow\ S^{'}_{quiet}(\lambda)$
				\State Weight $S^{'}_{quiet}(\lambda)$ by limb-darkening intensity $\mathbf{I}_{X,Y}$ and integrate spectra along disk surface. $\mathbf{S}_{quiet}+=\mathbf{I}_{X,Y} \times S^{'}_{quiet}(\lambda)$
			\EndFor

		\EndFor
	\end{algorithmic} 
\end{algorithm}

    The next step consists in initializing the active regions using the following parameters: the number of active regions, their size, their corresponding latitudes and longitudes, their types (either spot or faculae) and the resolution of the active region contour. An active region spectrum is also needed at this step to model the CB effect. The original SOAP 2.0 code uses the CCF of the observed spot spectrum in the visible obtained from the Kitt Peak Observatory FTS \citep[$S_{active} (\lambda)$ with $\lambda$ the same as for $S_{quiet} (\lambda)$][]{Wallace-2005}. The spectrum used for faculae regions is the same, with the difference that the contrast of such a region follow what is observed in the Sun \citep[e.g. Fig. 3 in][]{Meunier-2010a}, thus brighter than the quiet sun and with a center-to-limb brightening.
    Other groups use synthetic spectra at different temperature to model the quiet photospehere, spots and faculae, and include the effect of CB using results from magneto-hydrodynamical simulations \citep[e.g. the STARSIM 2 code\footnote{code available here \url{https://github.com/rosich/starsim-2}}][]{Herrero:2016aa}. We note that injecting observed or synthetic spectra have their advantages and drawbacks. Using observed spectra allows to better model the inhibition of convection inside active regions, but we note that if we use the observed spectra of a spot to model a facula (because an observed spectrum of a facula across the entire visible spectral range does not seem to exist), molecular features will be present in the facula spectrum despite the temperature being higher. Using synthetic spectra on the contrary allows to better model the temperature, and therefore spectral features, of the injected spectra. The choice of the spectra will be addressed in the later sections.
    
    In order to simulate a spectral time series, we need to calculate the disk location of active regions at each timestamp. As shown in Equation 1 and Equation 2 of \citep[][]{Boisse-2012b}, active regions are first put in the center of the disk and their initial configuration is obtained using a rotation matrix.
    Next, to get the position of those active regions as a function of time, another rotation matrix is used. At each timestamp, the code evaluates which active regions are visible, and which ones are hidden behind the star. This is performed by the function $\textbf{Localize}(lat, long, i, ph)$, where $lat$ and $long$ are the latitude and longitude of the active region center, $i$ is the inclination angle of the stellar disk and $ph$ is the rotational phase. The output of this function is a binary; one if visible, zero otherwise. If a region is visible, the code proceed to estimate the difference between the quiet solar spectrum and active spectrum at the location of the active regions. 
    
    The difference for the flux effect, the CB effect and the combination of the two (total) in each cell can be calculated using the following equations:
      \begin{equation} \label{eq:1}
      \Delta \mathbf{S}^{'}_{flux} (X,Y) = \mathbf{S}^{'}_{quiet} (X,Y) - \mathbf{I}_{ratio} \times \mathbf{S}^{'}_{quiet} (X,Y),
      \end{equation}
      \begin{equation}    \label{eq:2}   
      \Delta \mathbf{S}^{'}_{bconv} (X,Y) = \mathbf{S}^{'}_{quiet} (X,Y) - \mathbf{S}^{'}_{active} (X,Y),
       \end{equation}     
        \begin{equation} \label{eq:3}
            \Delta \mathbf{S}^{'}_{tot} (X,Y) = \mathbf{S}^{'}_{quiet} (X,Y) - \mathbf{I}_{ratio} \times \mathbf{S}^{'}_{active} (X,Y),
       \end{equation}
    where $\mathbf{I}_{ratio}$ is the contrast of the spot or faculae region. The code then integrates over all the cells covered by active regions to get final difference between the quiet spectrum and active spectrum. The final spectrum of each effect at each timestamp can be calculated by:
           \begin{equation} \label{eq:4}
            \mathbf{S}_{integrated, final} = \mathbf{S}_{integrated, quiet} - \Delta \mathbf{S}_{integrated, quiet-active}.
       \end{equation}  
       Once a spectrum for each timestamp is obtained, the code then lowers the resolution of the integrated spectrum to match the resolution provided in the configuration file. The pseudo code that describes how active regions are included, and how the final integrated spectra is obtained is summarized below.
  \begin{algorithm}
	\caption{Active region updates} 
	\begin{algorithmic}[1]
		 \For {$n_{region}=1,2,\ldots N$}
            \For {$t_{time step}=1,2,\ldots,T$}		 
                \State $\textbf{Localize}(lat, long, i, ph)$.
                \If {$\textbf{Localize} = 1$}
                    \State Shift $S_{quiet}(\lambda)$ with velocity $\mathbf{vel}_{X,Y}$. 
			        $S_{quiet}(\lambda)\rightarrow\ S_{quiet}(\lambda^{'})$.
			        \State Do linear interpolation to project the spectrum back to the original wavelength space.
			        $S_{quiet}(\lambda^{'})\rightarrow\ S^{'}_{quiet}(\lambda)$
				    \State Weight $S^{'}_{quiet}(\lambda)$ by limb-darkening intensity $\mathbf{I}_{X,Y}$

                   \State Shift $S_{active}(\lambda)$ with velocity $\mathbf{vel}_{X,Y}$. 
			        $S_{active}(\lambda)\rightarrow\ S_{active}(\lambda^{'})$.
			        \State Do linear interpolation to project the spectrum back to the original wavelength space.
			        $S_{active}(\lambda^{'})\rightarrow\ S^{'}_{active}(\lambda)$
				    \State Weight $S^{'}_{active}(\lambda)$ by limb-darkening intensity $\mathbf{I}_{X,Y}$
				    
				    \State Use Equations \ref{eq:1} to \ref{eq:3} to calculate the difference of each effect
				    
		        \EndIf
                \State Compute summation of $\Delta \mathbf{S}^{'} (X,Y)$ for each effect.
		
			\EndFor
		\EndFor

        \For {$t_{time step}=1,2,\ldots,T$}	
		 \For {$n_{region}=1,2,\ldots N$}
		    \State Use Equation \ref{eq:4} to update final spectrum at $t_{T}$.
            \State Lower the resolution of final spectrum at $t_{T}$ to match the HARPS-N observation.
        \EndFor
        \EndFor
	\end{algorithmic} 
\end{algorithm}

\subsection{The limitation of SOAP 2.0}    

The structure of SOAP 2.0 provides an efficient way to estimate stellar activities on spectroscopic measurement by simulating CCFs at different timestamps. The major drawback when changing the input from CCFs to spectra is the dimension of the data. The dimension of the input CCFs in SOAP 2.0 was $400$ in velocity space while the input high-resolution spectra we want to use have a dimension of $\sim 500000$ in the wavelength domain. From \textbf{Algorithm 1} and \textbf{Algorithm 2}, we clearly see that the linear interpolation is repeatedly called when injecting the spectrum in each cell, which is computationally expensive for an array with dimension of $\sim 500000$. For example, SOAP 2.0 takes $\sim 800$ seconds to calculate an integrated quiet sun spectrum using a $300 \times 300$ disk-grid. Another issue is how the code handles multiple active regions. Each active region is modeled independently, without information from other regions. This algorithm cannot handle the case in which some active regions overlap with each other. From real observations, we know that some active regions have complicated configurations. For example, most of active regions are a combination of a large faculae presenting a small spot in its center. In this context, a more computationally efficient and generalized algorithm is needed.

\section{Description of SOAP-GPU}\label{sec3}

In the previous section we've demonstrated the limitation of SOAP 2.0 when modeling stellar activity at the spectral level. Here, we present a new version of SOAP, based on Graphical Processing Unit (GPU) computing, that is much more efficient in term of computational speed, but also that adds some physical complexity.

\subsection{The basic concept of GPU computing}

The popularity of artificial intelligence has in recent year significantly increased due to the programmability of graphic hardwares. GPU computing uses graphical card as a co-processor for parallel computing. Compared with CPU, GPU solves problems by breaking them into separate tasks and processing them simultaneously. The basic computational unit that can independently perform simple calculation in a graphic card is called a $thread$. A group of threads that communicate and share memory with each other is called a $block$.

The new version of SOAP presented here, SOAP-GPU, is written using the Compute Unified Device Architecture (CUDA). CUDA is a compiler and toolkit for programming NVIDIA GPUs, and is an extension of the C/C++ programming language. CUDA invokes kernel functions by using the syntax of $<<<N_{blocks}, N_{threads} >>>$. This syntax allows the user to define the thread hierarchy before launching in parallel the same program function called $kernel$ to many threads. In order to launch the computation at the level of the GPU, a $host$ function defined in CPU controls the data transfer between CPU and GPU and can execute the kernel function inside the GPU. 

Threads in the same block can be accessed as 1D, 2D or 3D structures. In order to perform the thread level calculation, the index of individual thread and block need to be accessed. The index of each thread in the same block can be expressed as $threadIdx$. If the block is launched as the 1D structure, each thread in the same block can be accessed as $threadIdx.x$.
The number of the treads used in each 1D block can be obtained as $blockDim.x$. $Grid$ is a group of blocks. It can be either 1D, 2D or 3D. For the 1D grid, the index of each block in the grid can be expressed as $blockIdx.x$. Since the input spectra of quiet sun and active region are both 1D, we used the configuration of 1D grid with 1D block and the global index is  $index = blockIdx.x * blockDim.x + threadIdx.x$.
\subsection{Fast linear interpolation with GPU}
As mentioned in previous sections, the major limitation in SOAP 2.0 is the way it handles linear interpolation in each disk cell. A GPU provides thousands of cores which can be implement for linear interpolation for large data array. Considering that both quiet sun and spot spectra are evenly sampled in the wavelength domain, then the input wavelength can be described as:
\begin{equation} \label{eq:5}
    \lambda_{n} = \lambda_{0}+nk,
\end{equation}
where $n$ is the pixel number and $k$ is the step size. When a Doppler shift is applied, the wavelength array is modified as follow:
\begin{equation} \label{eq:6}
    \lambda^{'}_{n} = \lambda_{n}+\lambda_{n} f(\beta),    
\end{equation}
where $f(\beta) = -\left[1 - \sqrt{\frac{(1+\beta)}{(1-\beta)}}\right]$ and $\beta = v/c$. The variable $v$ is the velocity for each cell and $c$ is the speed of light.
Since we need to project the shifted spectrum back to the original wavelength space $S(\lambda^{'})\rightarrow\ S^{'}(\lambda)$, we have to find the index $m$ which satisfies $\lambda^{'}_{n} < \lambda_{m} < \lambda^{'}_{n+1}$.
For the left side, we have:
\begin{equation*} \label{eq:7}
    \lambda^{'}_{n} < \lambda_{m},
\end{equation*}
\begin{equation*}  \label{eq:8}
    \lambda^{'}_{n} = \lambda_{n}+\lambda_{n} f(\beta) = \lambda_{0}+nk + \lambda_{0} f(\beta)+nkf(\beta) < \lambda_{0}+mk,
\end{equation*}
so we have:
\begin{equation} \label{eq:9}
   n(1+f(\beta))+\frac{f(\beta)\lambda_{0}}{k} < m. 
\end{equation}
For the right side, we have:
\begin{equation} \label{eq:10}
   m < (n+1)(1+f(\beta))+\frac{f(\beta)\lambda_{0}}{k}. 
\end{equation}
Once the integer $m$ is known, we can estimate the flux for $\lambda_m$ using the spectrum derivative:

\begin{equation} \label{eq:11}
    S^{'}_{m} = \frac{\Delta S_{n}}{(\lambda^{'}_{n+1} - \lambda^{'}_{n})} \times (\lambda_{m} - \lambda^{'}_{n})+S_{n},
\end{equation}
where $S_{n} = S(\lambda_n)$ and $\Delta S_{n} = S_{n+1} - S_{n}$. 

Equations \ref{eq:9} to \ref{eq:11} can be parallelised using GPU. We launch 1D grid of 1D blocks with $<<<N_{blocks}, N_{threads} >>>$ to perform the linear interpolation mentioned above and the number of blocks and threads satisfies $Dim_{input\_spectrum} = N_{blocks} \times N_{threads}$. The pseudo code for this part is summarised in \textbf{Algorithm 3} and the quiet sun spectra integration can be rewritten as \textbf{Algorithm 4}.
\begin{algorithm}
	\caption{Fast interpolation with GPU} 
	\begin{algorithmic}[1]
	    \State $index = blockIdx.x * blockDim.x + threadIdx.x$
	    \State $index_{target} = ceil(index*(1+f(\beta))+f(\beta)*\lambda_{0}/k)$
	    \State $S^{'}_{index_{target}} = \frac{\Delta S_{index}}{(\lambda^{'}_{index+1} - \lambda^{'}_{index})} \times (\lambda_{index_{target}} - \lambda^{'}_{index})+S_{index}$.
	\end{algorithmic} 
\end{algorithm}

\begin{algorithm}
	\caption{Quiet spectrum integration with GPU} 
	\begin{algorithmic}[1]
		\For {$X_{location}=1,2,\ldots N$}
			\For {$Y_{location}=1,2,\ldots,N$}
			    \State Apply Doppler shift with velocity $\mathbf{vel}_{X,Y}$ and derive $S^{'}_{quiet}(\lambda)$ using GPU fast interpolation
				\State Weight $S^{'}_{quiet}(\lambda)$ by limb-darkening intensity $\mathbf{I}_{X,Y}$ and integrate spectra along disk surface. $\mathbf{S}_{quiet}+=\mathbf{I}_{X,Y} \times S^{'}_{quiet}(\lambda)$
			\EndFor
		\EndFor
	\end{algorithmic} 
\end{algorithm}

\subsection{Active region updates}

As addressed in the previous section, one of the disadvantage of SOAP 2.0 is that each active region is modeled independently, which makes the code unable to handle complicated active region configurations: some active regions may overlap with each other; spots may be surrounded by facualae regions. Here we propose a revised algorithm to update active regions: an empty disk map called $InfoMap$ is allocated in the GPU first. At each timestamp, A list of active regions with their properties is uploaded and the code calculates the location of active regions projected on the disk map. If some regions are visible, we update the corresponding pixels with their active region types in the information map. For example, if a faculae region is visible at ($x_n, y_n$), $InfoMap(x_n,y_n) = 1.0$. If there are multiple regions with the same type overlapping with each other, the overlapping region in the information map will remain the same. This will avoid the over-calculation for the overlapping region issue in the SOAP 2.0 since each acitve region is calculated independently. This algorithm can also simulate complicated active region configurations. For example, a spot surrounded by a large faculae can be simulated by updating the information map with a faculae first. If the spot region is embedded inside the faculae, the overlapping region in the information map will be updated with the type of the spot. The pseudo code of this part is summarised in \textbf{Algorithm 5}.

  \begin{algorithm}
	\caption{Active region updates with GPU} 
	\begin{algorithmic}[1]
		\For {$t_{time step}=1,2,\ldots,T$}
		    \For {$n_{region}=1,2,\ldots N$}
              \State $\textbf{Localize}(lat, long, i, ph)$.
              
                \If {$\textbf{Localize} = 1$}                
                
                    \State Updating $InfoMap(x_n, y_n) = $ the type of active region.
				    
		        \EndIf

			\EndFor

                \State Inject velocity $\mathbf{vel}_{X,Y}$ with GPU fast interpolation for the active regions in the information map.
                and derive $S^{'}_{active}(\lambda)$.
				\State Weight $S^{'}_{active}(\lambda)$ by limb-darkening intensity $\mathbf{I}_{X,Y}$
				\State Use Equations \ref{eq:1} to \ref{eq:3} to calculate the difference of each effect
                \State Compute summation of $\Delta \mathbf{S}^{'} (X,Y)$ for each effect.	
            \State Use Equation \ref{eq:4} to derive the final spectrum at $t_{T}$.
            \State Lower the resolution of final spectrum at $t_{T}$ to match the HARPS-N observation.
		\EndFor
	\end{algorithmic} 
\end{algorithm}

\subsection{Differential rotation}

In the original SOAP 2.0 code, there is no differential rotation implemented. In order to better model stellar activity, differential rotation is included when the stellar disk is initialized according to the equation $\omega = \omega_{0}+\omega_{1}\sin^{2}(\theta)$, where $\omega_{0} = 14.371^{\circ}/\rm{day}$ and $\omega_{1} = -2.587^{\circ}/\rm{day}$ for the Sun \citep[][]{Borgniet-2015aa}. To generalise this for other stars, the user can select in the configuration of SOAP-GPU a rotation period and a differential rotation rate. $\omega_{0}$ is then equal to 360/PROT and $\omega_{1}$ to DIFF\_ROT*PROT (PROT=25.05 and DIFF\_ROT=-0.18 for the solar case to reproduce the above equation).

\section{Results}\label{sec4}
  \begin{figure}[htbp]
  \includegraphics[scale=0.35]{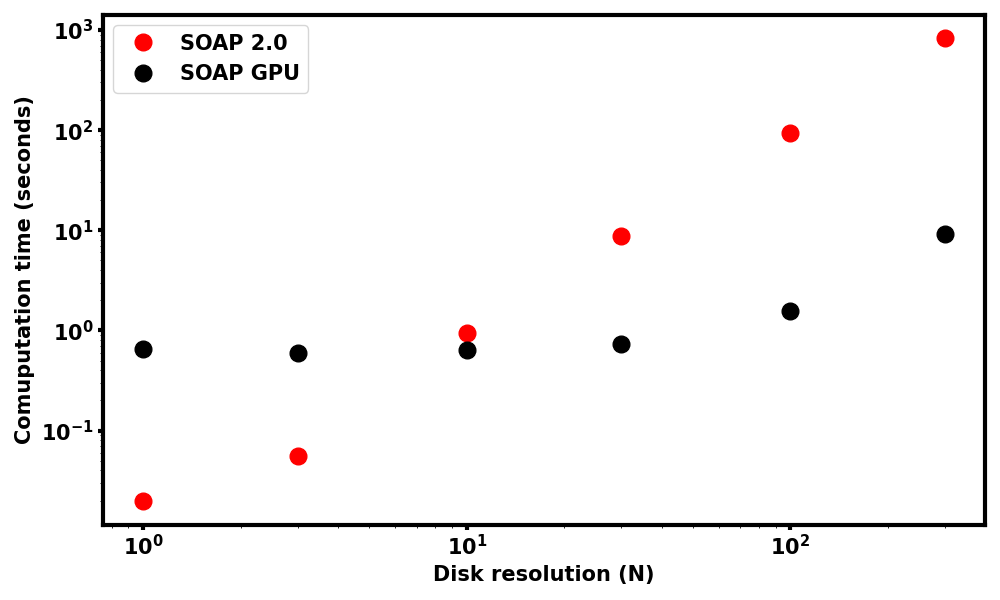}
  \caption{The computation speed comparison between SOAP 2.0 and SOAP GPU: The integrated quiet sun spectrum is calculated with different disk resolutions. When disk resolution is below 10, SOAP 2.0 is faster than SOAP-GPU since the communication between CPU and GPU in SOAP-GPU is time-consuming. For resolutions above 10, SOAP-GPU is significantly faster than SOAP 2.0. With a typical resolution value of 300, the quiet disk spectrum integration in SOAP-GPU is 100 times faster than in SOAP 2.0}
    \label{Fig1}%
    \end{figure}

\subsection{Performance and precision comparison with SOAP 2.0}

We examined the performance of SOAP-GPU in two aspects: computational speed and accuracy. SOAP-GPU code is executed on a Nvidia RTX-3090 card while we run the modified SOAP 2.0 that generates spectra in a MacBook Pro with 2.6 GHz 6-Core Intel Core i7. We analysed the speed performance of SOAP-GPU by calculating the time it takes to obtain an integrated quiet sun spectrum. The input quiet sun spectrum has a dimension of 547840, thus the kernel function fast interpolation is launched with $<<<N_{blocks}, N_{threads} >>> = <<<1070, 512 >>>$. We note that $N_{threads}$ is fixed to 512 and $N_{blocks}$ is an adaptive number based on the dimension of the input. SOAP 2.0 is executed with the same simulation configuration on a single CPU. We computed the integrated quiet sun spectrum with different disk resolution and their computational time is show in Figure~\ref{Fig1}. When the disk resolution is very low, smaller than 10, SOAP 2.0 is faster than SOAP-GPU. This is not surprising since the data transfer between GPU and CPU in SOAP-GPU is the dominating factor. When the disk resolution increases, SOAP-GPU is significantly faster than SOAP 2.0. When the resolution is above 100, the quiet sun spectrum integration of SOAP-GPU is 100 times faster than SOAP 2.0 and both computational curves linearly increase in log-log space.

\cite{Boisse-2012b} found no significant change in their
results with resolution beyond 300, therefore, we used this disk resolution for the following of the paper. For the typical disk resolution of 300, a spot at disc center with an area of $1\%$ of the entire disk will be contained in a grid of $34 \times 34$ cells. Due to the small size of the grid for such a configuration, the fast interpolation algorithm (see \textbf{Algorithm 3}) is only able to gain a factor of $\sim$10 in computation time. If the spot size increases to $9 \%$ of the entire disk, the simulation can then gain almost the full speed boost from fast interpolation (100 time faster). Fast interpolation at the level of the active region modelisation makes therefore significant improvements in computational speed when considering high-resolution simulations or simulations with large active regions.

 \begin{figure}[htbp]
  \includegraphics[scale=0.35]{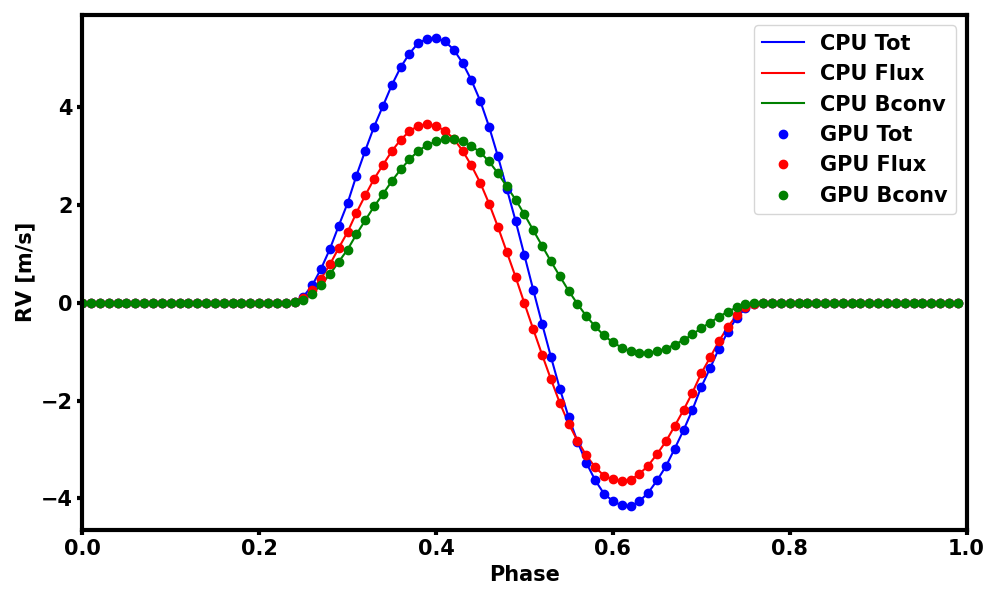}
  \caption{Comparison of the RVs derived from the simulated spectra modeled by SOAP2.0 and SOAP-GPU. A single equatorial spot with 1\% area of the entire disk surface is simulated. It took 1749.3 seconds to simulate those spectra with SOAP 2.0 while only 27.9 seconds with SOAP-GPU on a Nvidia RTX-3090 card. The computation speed is improved by a factor of 63.}
    \label{Fig2}%
    \end{figure}

We also examined the accuracy of SOAP-GPU.  We simulated a single equatorial spot with a $1\%$ area of the entire disk surface using a disk resolution of 300. It took 1749.3 seconds to simulate those spectra with SOAP 2.0 for 100 timestamps while only 27.9 seconds using SOAP-GPU, which corresponds to a gain of a factor 63. The modeled RVs relative to the flux effect, the CB effect and the total effect are derived from the simulated spectra by cross-correlating them with the same mask originally used in SOAP 2.0, and measuring the RV as the mean of a Gaussian profile fitted to the obtained CCFs. Figure~\ref{Fig2} illustrates that the simulated spectra from SOAP-GPU provides the same RVs as the spectra from SOAP 2.0.

\subsection{Exploration of active region properties}

The dynamics of active regions plays an important role for understanding the stellar activity-induced RVs. Most of previous study aimed at investigating these effects with real observations. For example, \citet{Meunier-2010a} derived the stellar activity induced RVs by using Michelson Doppler Imager/Solar and Heliospheric Observatory (MDI/SOHO) magnetograms images. At the simulation level, \cite{Gilbertson:2020aa} investigated the effect of spot evolution on the long-term and at the spectral level, using a modified version of SOAP 2.0. However, they only considered spots, and only their decaying phase. In order to illustrate the effects of active region dynamics, we discuss in this section the photometric and RV variations observed when an active region changes in size, when different number of active regions are present and when the active region configuration changes. 

\subsubsection{The size evolution of active region}

\begin{figure*}
  \centering
  \includegraphics[scale=0.45]{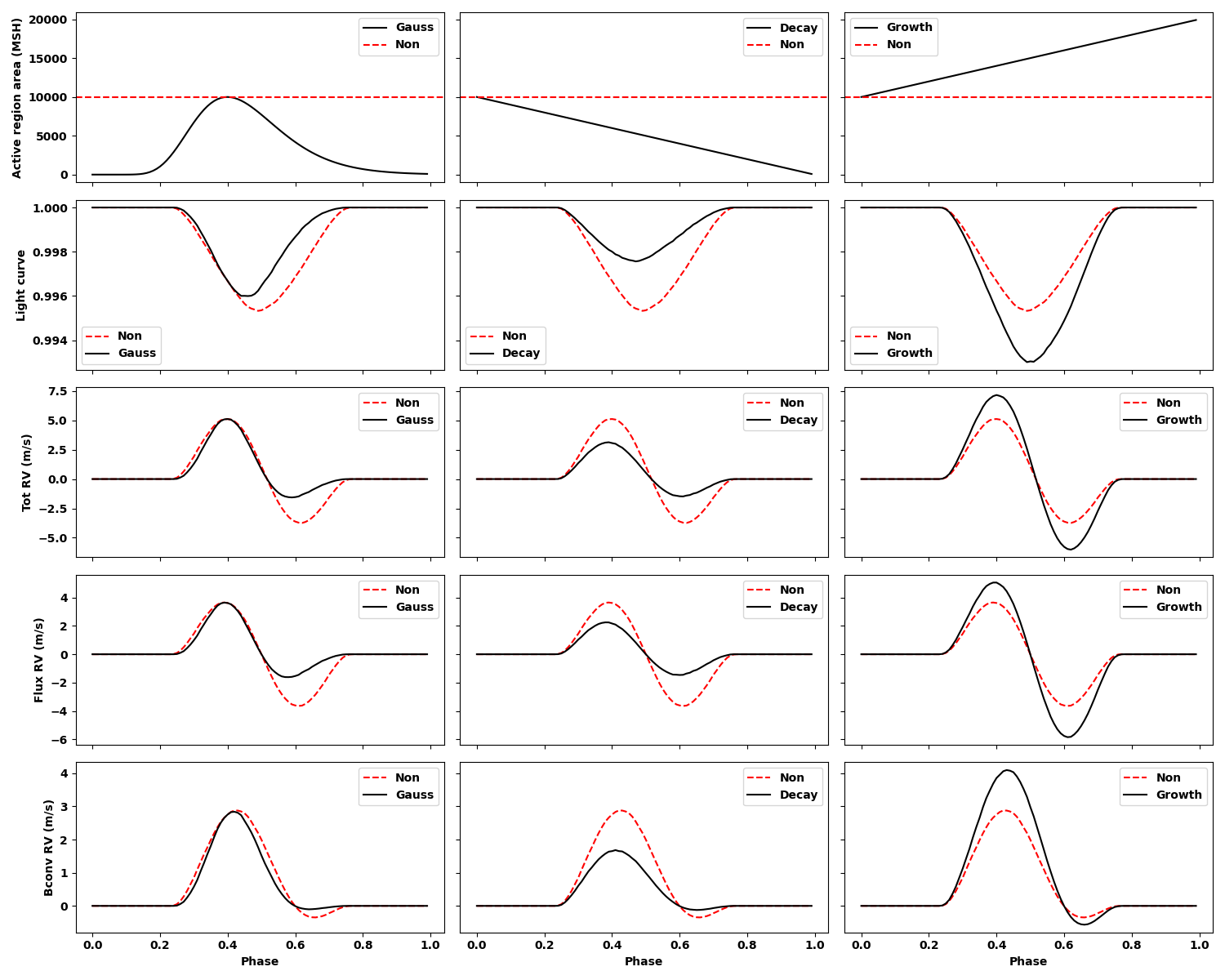}
  \caption{SOAP-GPU simulation of different active region evolution curves: A single spot with a latitude of 30\degree and longitude of 180\degree is simulated. Three spot size evolution types are demonstrated: i) a fast growth and slow decay evolution is shown in the first column, ii) a linear decay evolution curve in the second column and iii) a linear growth curve in the third column. The evolution curves and the simulated light curves are shown in the first two rows. The RVs of the total effect, the flux effect and the CB effect are present in the rest of the rows. The simulation of an non-evolving spot (red dashed line ) is also shown in each figure for comparison.}
    \label{Fig5}%
    \end{figure*}

To explore different active region evolution scenarios, we developed and included an evolution module in SOAP-GPU. This module can model evolution in three different ways: i) a linear growing phase, ii) a linear decaying phase or iii) a growing and decaying phase modeled by an an asymmetric Gaussian function \citep{Murakozy:2014SoPh}. Other user-defined functions can be added to this module if desired. We show in Fig.~\ref{Fig5} the impact of active region evolution on the light-curve and on the different RVs derived (flux, CB and total effects). For the asymmetric Gaussian evolution phase, the maximum size is set to 10000 millionths of solar hemisphere (MSH) equivalent to 1\% of the visible hemisphere, the FWHM to 10 days and an asymmetry factor of 0.09. For the growing only, or decaying only evolution phases, the initial size is set to 10000 MSH and the growth or decay rate is set to 400 MSH/day. We found that both flux and CB effects are sensitive to the evolution of active regions.

\subsubsection{Complex active regions}

\begin{figure}[htbp]
  \includegraphics[scale=0.35]{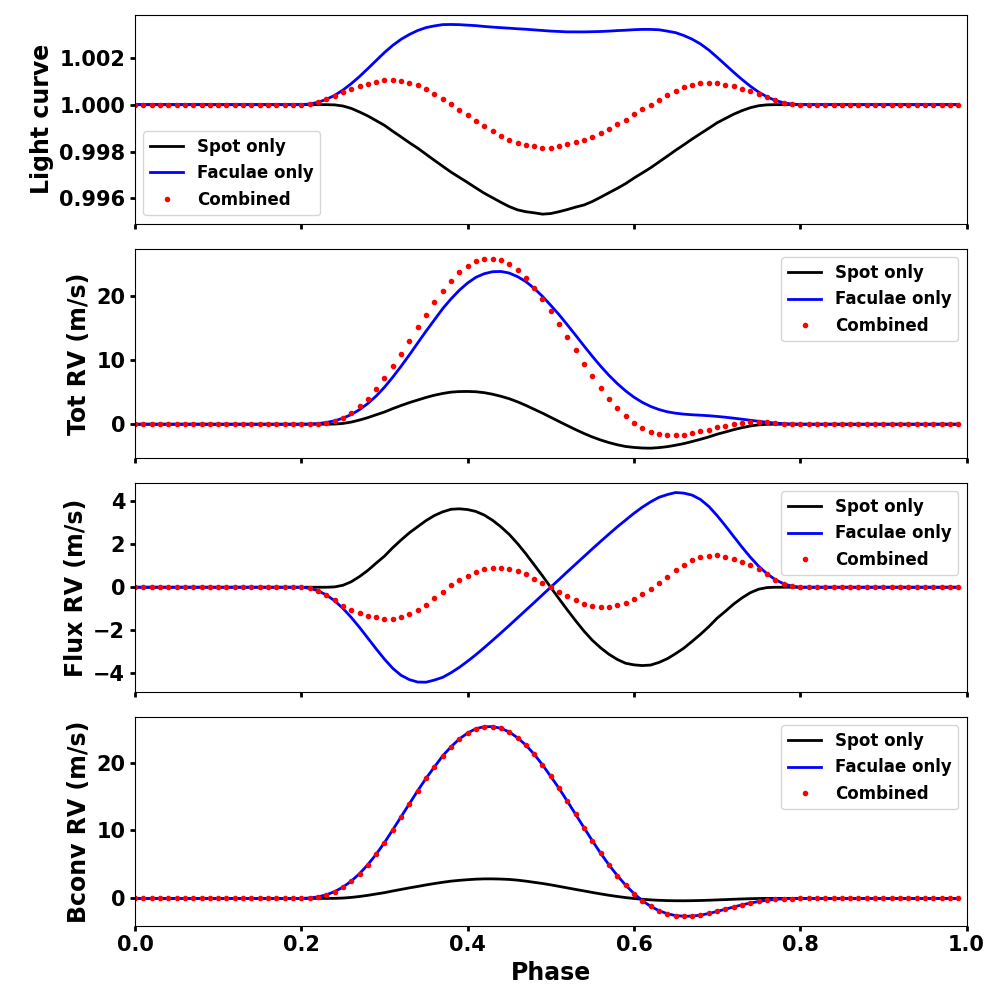}
  \caption{Three different active region configurations are simulated. A single spot located at the latitude of 30\degree and the longitude of 180\degree with a fixed size of $1\%$ of the entire solar disk is present in black. A single facula with the same coordinates and a fixed size of $9\%$ of the entire solar disk is shown in blue line. The $1\%$ spot region surrounded with $9\%$ faclua region is labeled in red dashed line. The top panel demonstrates the light curves of different configurations and the rest of the panels shows the RVs of the total effect, the flux effect and the CB effect.}
    \label{Fig6}%
    \end{figure}

SOAP-GPU also allows users to simulate complex active region configurations. From the observational point of view, facluae and spots are not independent from each other. The facula distribution is based on the spot distribution. This leads to a complex configuration in which spots may overlap faculae \citep[][]{Borgniet-2015aa, Chapman-2001ApJ}. In order to model such a configuration, the SOAP-GPU config file allows users to define the distribution of active regions, as a sequence of spots and faculae with given properties (size, initial longitude, initial latitude).
For example, in order to simulate a spot surrounded by a facula, the user can define the location and the size of the large facula first and then define a smaller spot at the same location. The region of overlap will be replaced by the spot as mentioned in \textbf{Algorithm 5}. A simulation of this case is illustrated in Figure ~\ref{Fig6}, with a $1\%$ spot surrounded by a $9\%$ facula. Since the spot has a higher contrast than the faculae, the light curve and RVs of the flux effect is dominated by the spot while the RVs of the CB effect is dominated by facula. Overall, the CB RV effect induced by the facula dominates all the other contributions, and thus the total RVs is affected mainly by the facula, as it was already demonstrated in several studies \citep[e.g.][]{Meunier-2010a,Dumusque-2014b,Milbourne:2019ApJ}.

\subsection{Exploration of spectral properties}

In this section, we explore the input spectra properties and demonstrate how the derived RV behaves depending on the wavelength domain.

\subsubsection{Chromatic effects of different wavelength coverage} \label{chromatic_effect}

\begin{figure}[htbp]
  \centering
  \includegraphics[scale=0.35]{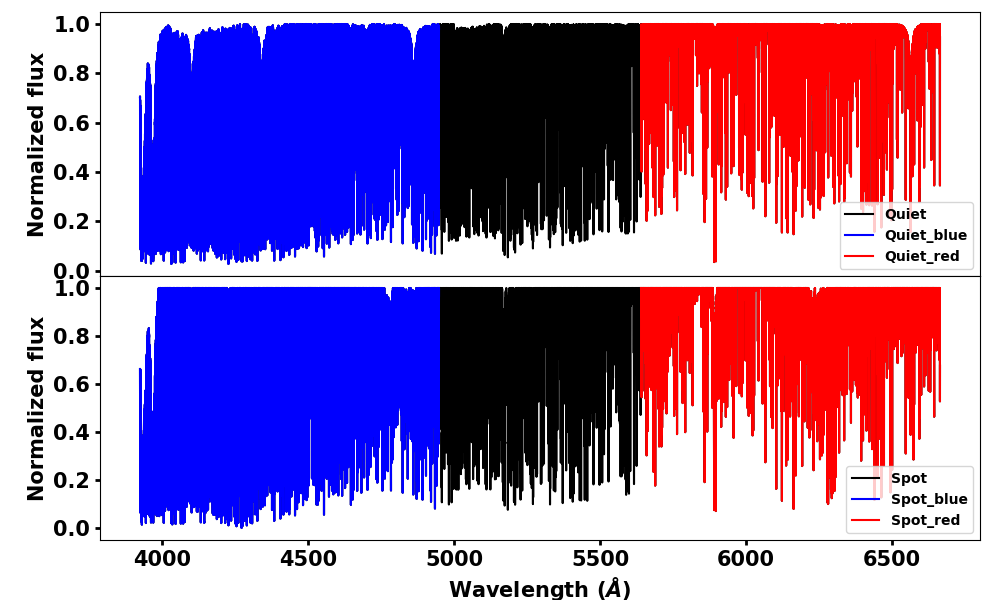}
  \caption{Injecting different size of spectra in SOAP-GPU input. Three different sets of quiet sun and spot spectra, with different wavelength ranges are used as input to SOAP-GPU: The entire spectra with length 547840 is labeled in black. The blue and red parts of the spectra with length 204800 are over plotted in blue and red, respectively.}
    \label{Fig3}%
    \end{figure}

To explore the effect induced by different wavelength coverage, we injected into SOAP-GPU only the red or only the blue part of the quiet sun and spot spectra (see Fig.~\ref{Fig3}). The red and blue parts have the same dimension of 204800, which is different from the full spectra. As the fast interpolation kernel function depends on the dimensions of the input spectra, the code automatically configures the kernel with the option $<<<400, 512 >>>$.

  \begin{figure}[htbp]
  \centering
  \includegraphics[scale=0.35]{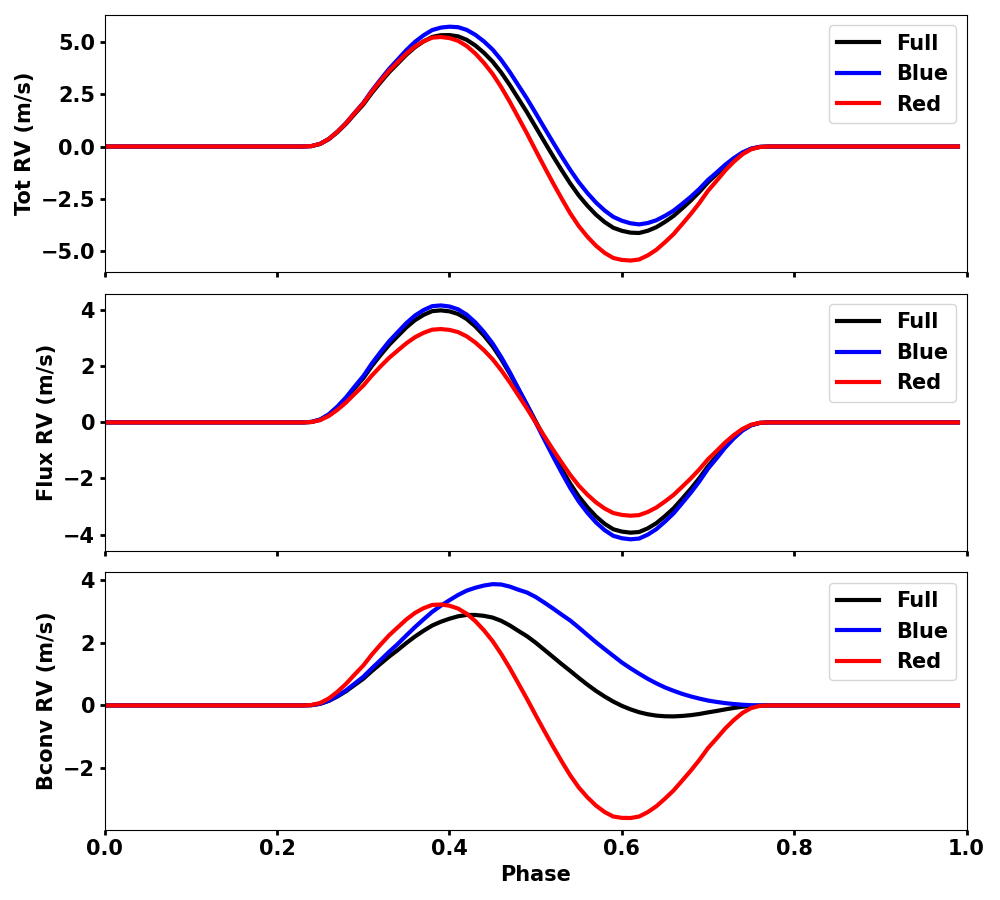}
  \caption{The chromatic effect of RVs. SOAP-GPU is initialized with three different spectra: entire wavelength coverage, and only red and blue spectral parts (see Sect.~\ref{Fig3}). A single spot with a size of 1\%, a latitude of 30\degree and a longitude of 180\degree is modeled by the code. The measured RVs with different input spectra are labeled with black, red and blue, respectively. Top: The measured RVs of the total effect. Middle: The measured RVs of the flux effect. An offset of 1 $\rm{m s}^{-1}$ is added in the red and blue RVs. Bottom: The measured RVs of the CB effect.}
    \label{Fig4}%
    \end{figure}

The measured RVs are shown in Figure~\ref{Fig4}. The RVs of the CB effect are different between the blue and red parts. One notable thing is that the RVs of the CB effect simulated from the red inputs goes below zero, while we expect the CB effect to only be positive, as it corresponds to an inhibition of CB. To confirm that nothing was wrong at the level of the code, we injected for the spotted region the same spectrum as the quiet Sun, but we red-shifted it by 300 m/s to model at first order the inhibition of CB. In this idealist case, the CB effect does not provide negative values. After further investigation, those negative values comes from the fact that the spot temperature is lower than the quiet photosphere. Thus, spectral lines will change in depth, which will induce a flux effect even when not considering the contrast of the active regions when estimating the CB effect (see Eq.~\ref{eq:2}). In the case of the Sun, this flux effect seen in the CB derived RVs is mainly coming from the red part of the spectrum due to molecular absorption that can be seen in the spot spectrum, but not in the quiet photosphere spectrum. As we will
see in Sect.~\ref{phoenix_spectra}, we do not obtain negative values when injecting PHOENIX solar equivalent spectra for the quiet and active Sun instead of the Kitt Peak solar quiet and active atlases, however we still see a slight asymmetry in the derived CB RV effect, pointing toward a small flux effect contribution. This is likely because PHOENIX spectra are not able to model all the absorptions coming from molecular bands, and thus the flux effect seen in the CB estimation only is stronger for the real solar spectra than for the synthetic spectra. 
This issue prevent us of fully separating the flux from the CB effect, however, we note that the total effect (flux + CB) should be modeled properly. We note that this feature was not visible in SOAP 2.0, as after computing the CCF for the quiet and actives regions, we were renormalising them.

We note that in SOAP 2.0, we used a fixed contrast to model the flux effect of active regions. This contrast was derived by comparing the Planck function of the quiet Sun effective temperature and of the spot or facula temperature\footnote{the config file in SOAP 2.0 allows to give an effective temperature for the quiet photosphere, 5778 $K$ for the Sun, and a temperature difference with respect to this former value for the spot spectrum (663 $K$ as the default value in SOAP 2.0). For a facula, the temperature is dependent on the center-to-limb angle and was following what is observed on the Sun \citep[e.g. Fig.~3 in][]{Meunier-2010a}.}, at the average wavelength of the input spectra 5293 $\AA$. Now that we use spectra as input, and not CCFs, we implemented a contrast that is wavelength dependant to model the chromatic effects of stellar activity. To do so, we introduced a new GPU kernel function called SOAP$contrast\,<<<N_{blocks}, N_{threads} >>>$. This new kernel allows to perform the wavelength dependent contrast calculation: each wavelength pixel is first accessed by the global index of the kernel. Next on each thread, it derives the contrast by calculating the ratio of two Planck functions at two different effective temperatures. The absolute value of the contrast in the blue part of the spectrum is higher than in the red part. This implies that the flux effect for the blue part is stronger than in the red part, which can be seen in the middle panel of Fig.~\ref{Fig4}.

\subsubsection{Convection as a function of center-to-limb angle} \label{CB_for_mu}

Solar spectral line profiles become asymmetric due to convective motions varying with physical depth inside the solar photosphere \citep[e.g.][]{Dravins-1981, Gray-2009}. This effect also leads to a change in shape of the bisector of spectral lines from disk-center to the limb, as photons are coming from different physical depths \citep[e.g.][]{Cavallini-1985b}. In order to better model the effect of convection in SOAP-GPU, we derived this effect from very-high spatial and spectral resolution observations of the Sun \citep{Lohner-Bottcher:2018aa,Stief:2019aa,Lohner-Bottcher:2019aa}.

\begin{figure*}[htbp]
  \centering
  \includegraphics[scale=0.45]{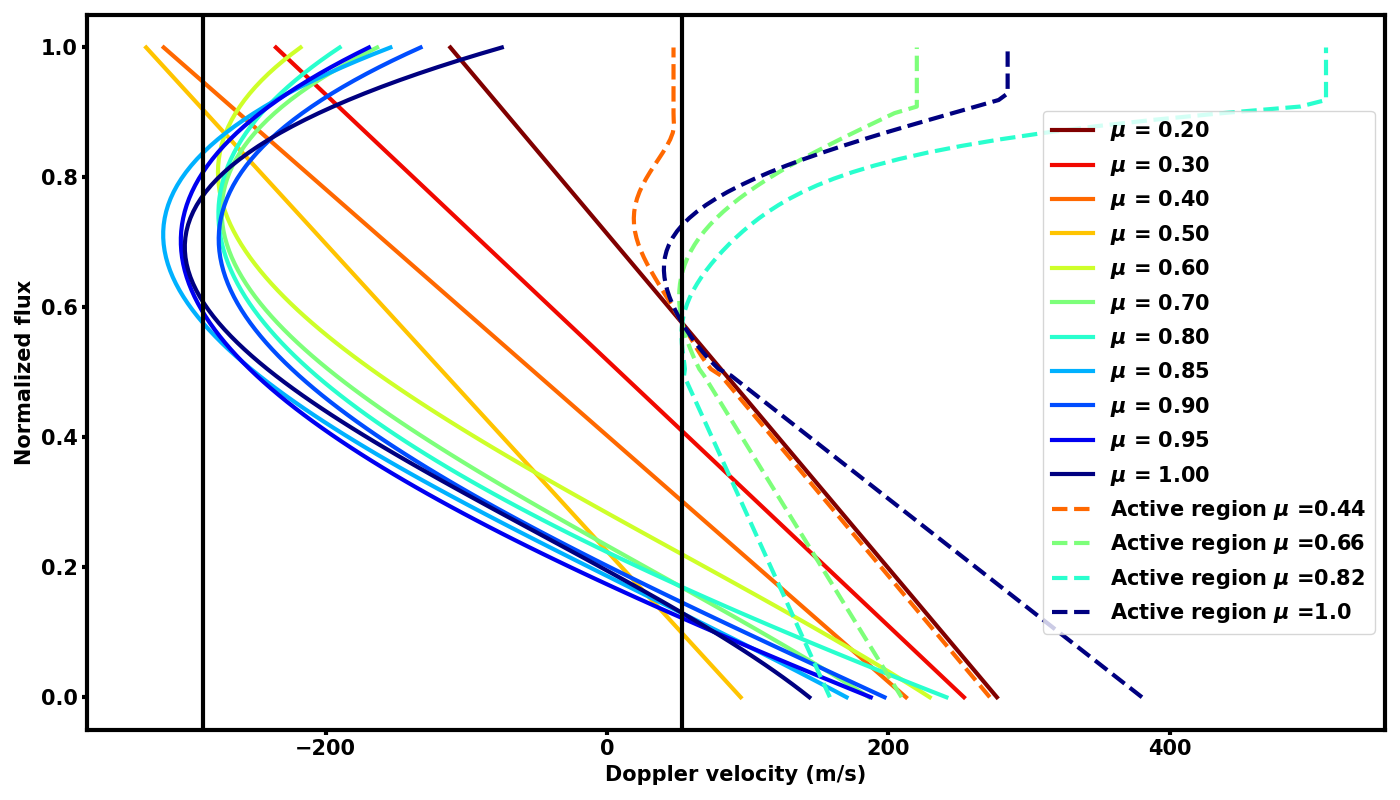}
  \caption{Average bisectors of quiet and active solar regions from the disk center ($\mu = 1.0$) to the limb ($\mu = 0.2$). \emph{Continuous lines:} Fifth-order polynomial fit to the quiet sun bisectors of the FeI $5250.2084\AA$, FeI $5250.6453 \AA$, FeI $5434.5232\AA$, FeI $5432.9470\AA$, FeI $5576.0881\AA$, FeI $6149.2460\AA$, FeI $6173.3344\AA$, FeI $6301.5008\AA$ and FeI $6302.4932\AA$ lines as measured by the Laser Absolute Reference Spectrograph (LARS) at the German Vacuum Tower Telescope \citep[][]{Lohner-Bottcher:2019aa}. \emph{Dashed lines:} Fit of the bisectors of the FeI $6301.5008\AA$ spectral line inside a faculae region, as measured by the Fabry-Perot interferometer at the Donati Solar Tower \citep[][]{Cavallini-1985}. Below a depth of 0.5, a linear fit is performed, while a fifht-order polynomial is used to model the top part of the bisector. To prevent unrealistic value when interpolating the polynomial above a normalised flux of 0.9 where no measurement exists, we selected the most redshifted part of the top bisector, explaining the vertical values for very shallow depths. The two vertical lines are shifted by 340 m/s which corresponds to the solar convective blueshift value derived from a fit to the data of \citet{Liebing:2021aa} (see Sect.~\ref{phoenix_spectra}). The active bisectors at different $\mu$ angles are all shifted by those 340 m/s at a depth of 0.58 as we make the hypothesis that convection is fully suppressed in magnetic regions (see Sect \ref{CB_for_mu} for more information).}
    \label{Fig_mu_BIS}%
    \end{figure*}

We note that the varying shape of spectral line with center-to-limb $\mu$ angle is also modelled in the STARSIM 2 code \cite[][]{Herrero:2016aa} by fitting a fourth-order polynomial function on magneto-hydrodynamic CIFIST 3D models \citep{Ludwig-2009MmSAI}. However, this fifth-order polynomial is only valid for line depth as strong as $\sim 0.5$ \citep[see Fig.~6 in][]{Herrero:2016aa}\footnote{The parameters mentioned here are derived from the published code ( \url{https://github.com/rosich/starsim-2})}. This was enough to model the shape change of the CCFs in STARSIM 2, which does not go deeper. However, when working at the spectral line level, this polynomial will give completely wrong estimate for the core of deep lines, due to extrapolation. We therefore used the quiet sun observations at different $\mu$ angles provided in \citet{Lohner-Bottcher:2019aa}. We first measured the bisectors of all the available iron deep lines at different $\mu$ angle in \citet{Lohner-Bottcher:2019aa}\footnote{We use the following lines: FeI $5250.2084\AA$, FeI $5250.6453 \AA$, FeI $5434.5232\AA$, FeI $5432.9470\AA$, FeI $5576.0881\AA$, FeII $6149.2460\AA$, FeI $6173.3344\AA$, FeI $6301.5008\AA$ and FeI $6302.4932\AA$} 
, and fitted them using polynomial functions. For $\mu$ angle smaller then 0.5, we used a straight line to fit the bisectors of the selected deep lines, to prevent strong divergence when extrapolating the fit towards very large depths. For $\mu$ angle larger or equal to 0.5, 3rd order polynomial functions are used to capture the curvature of the bisectors around disk center. The final bisectors, shown in Fig.~\ref{Fig_mu_BIS} are obtained by interpolating and extrapolating those bisectors from depth 0 to 1.

To obtain the dependency of the spectral line bisector as a function of $\mu$ in an active region, we use the observations presented in \citet{Cavallini-1985}. We parameterised the bisectors of the FeI at $6301.5008\AA$, for the disk center ($\mu=1$) and different center-to-limb angles ($\mu=0.82$, 0.66 and 0.44). The bottom part of the bisectors, below 0.5, is fitted using a straight line, the upper part for which we have data, using a 5th-order polynomial. Rather than extrapolating the fitted polynomial towards very shallow depths, which can give unrealistic redshifted values, we decided to use the more redshifted data value of the top bisector for extrapolation. We show in Fig.~\ref{Fig_mu_BIS} the obtained active bisectors from depth 0.0 to 1.0.

\begin{figure*}[htbp]
    \centering
  \includegraphics[scale=0.40]{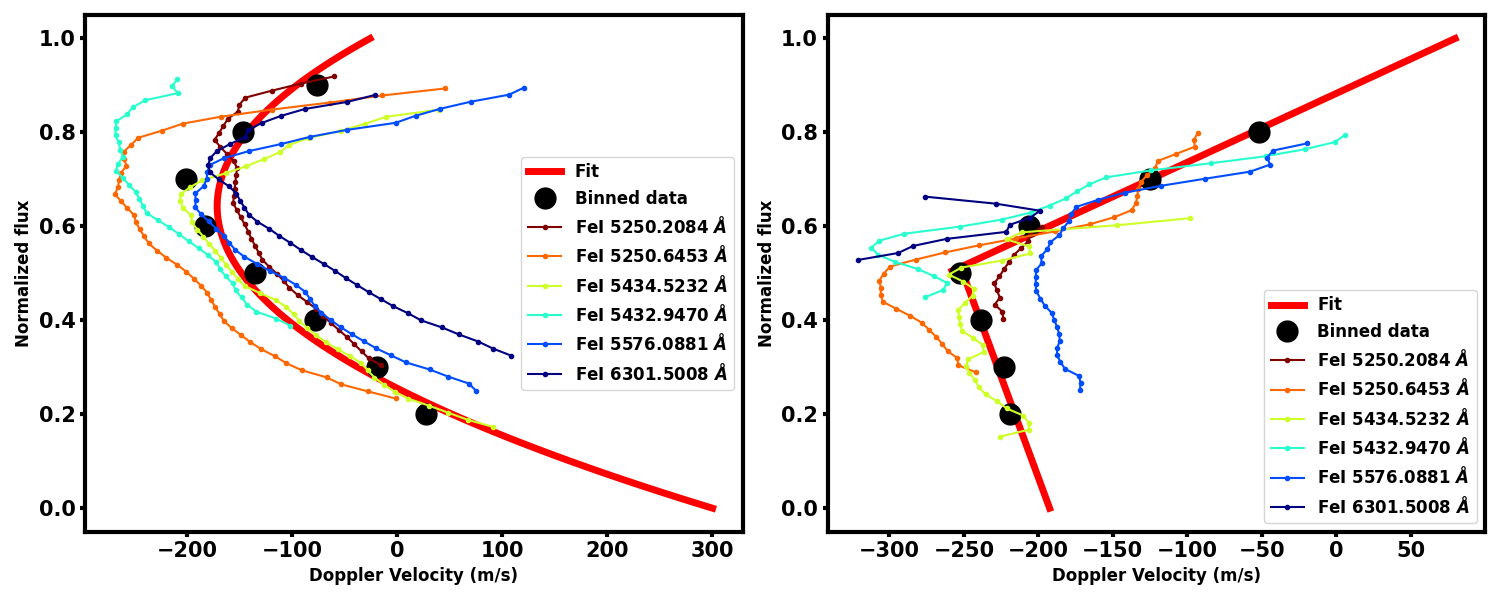}
  \caption{Bisectors of the FeI lines used in \cite{Lohner-Bottcher:2019aa} and fitted model to account for the CB. \emph{Left:} Bisectors from the quiet Kitt Peak solar spectrum. \emph{Right:} Bisectors from the active Kitt Peak solar spectrum. We rejected the bottom part of the $6301.5008\AA$ bisector because it was significantly off by 2500 m/s due to  strong contamination by other weak lines. }
    \label{Fig_CB_inhibition_kitt_peak}%
    \end{figure*}

Once we have our model for line bisectors at different $\mu$ angles, we can use the Python module $Convec.model$ to apply those bisectors to the original spectra, and thus obtain different spectra for different $\mu$ angles. Each cell in the stellar disk takes the bisector that has the closest $\mu$ angle. However, the code first has to remove the original bisector from the spectral lines of the input quiet and active Kitt Peak solar spectra. To do so, we select the same lines as in \citet{Lohner-Bottcher:2019aa} in the input spectra and measure their individual bisectors. To model the average bisector of the lines selected in the quiet spectrum, we use a second-order polynomial. For the active spectrum, due to the lower effective temperature, the wings of certain lines fitted are blended, which strongly impact the bisector measurement. We therefore rejected bisector points that are significantly off.  Then, we model the average active bisector of the lines by fitting the regions below and above a depth of 0.5 with two different linear models. Fitting a higher-order polynomial for those active bisectors was giving unrealistic values when extrapolating to very small or very large depths. The measured individual bisectors with our models are shown in Fig.~\ref{Fig_CB_inhibition_kitt_peak}. To finally obtain quiet and active spectra with proper bisector shape as a function of $\mu$ angles, we remove the original bisectors of the quiet and active Kitt peak solar spectra, and then add the bisectors measured for different $\mu$ angles. This is done by shifting each point in those spectra depending on their normalised depth.

To inject the proper bisectors for different $\mu$ angles in our original spectra, we first remove the original bisector, which changes any CB difference between the quiet and active solar spectra. We therefore need to impose a shift between the bisectors of quiet and active regions in order to properly model the inhibition of CB inside active regions. We here make two assumptions: i) the CB is fully inhibited at $\mu$=0.2 in the quiet Sun, and ii) it is also fully inhibited for magnetic regions, and this at all $\mu$ angles. Using the first assumption, we measure for the quiet Sun the maximum shift between the bisector at $\mu$=0.85, which is the bisector that is the most blueshifted, and the bisector at $\mu$=0.2. This maximum happens at a depth of 0.58 and equals to 375 m/s. This value is extremely similar to the 340 m/s CB value derived from a fit to the data of \citet{Liebing:2021aa} (see Sect.~\ref{phoenix_spectra}). To match the CB relation derived in Sect.~\ref{phoenix_spectra}, we rescale the maximum difference between the $\mu=0.85$ and $\mu=0.2$ to be 340 m/s. We note that this rescaling is negligible in the case of the Sun, however, it will be really needed in Sect.~\ref{phoenix_spectra} when using PHOENIX spectra as input. Using the second assumption, we impose that at the same depth of 0.58, the difference in velocity between the quiet bisector at $\mu$=0.85 and all active bisectors is also 340 m/s. We show the proper shift between the quiet and active bisectors in Fig.~\ref{Fig_mu_BIS}.

\begin{figure}[htbp]
  \includegraphics[scale=0.35]{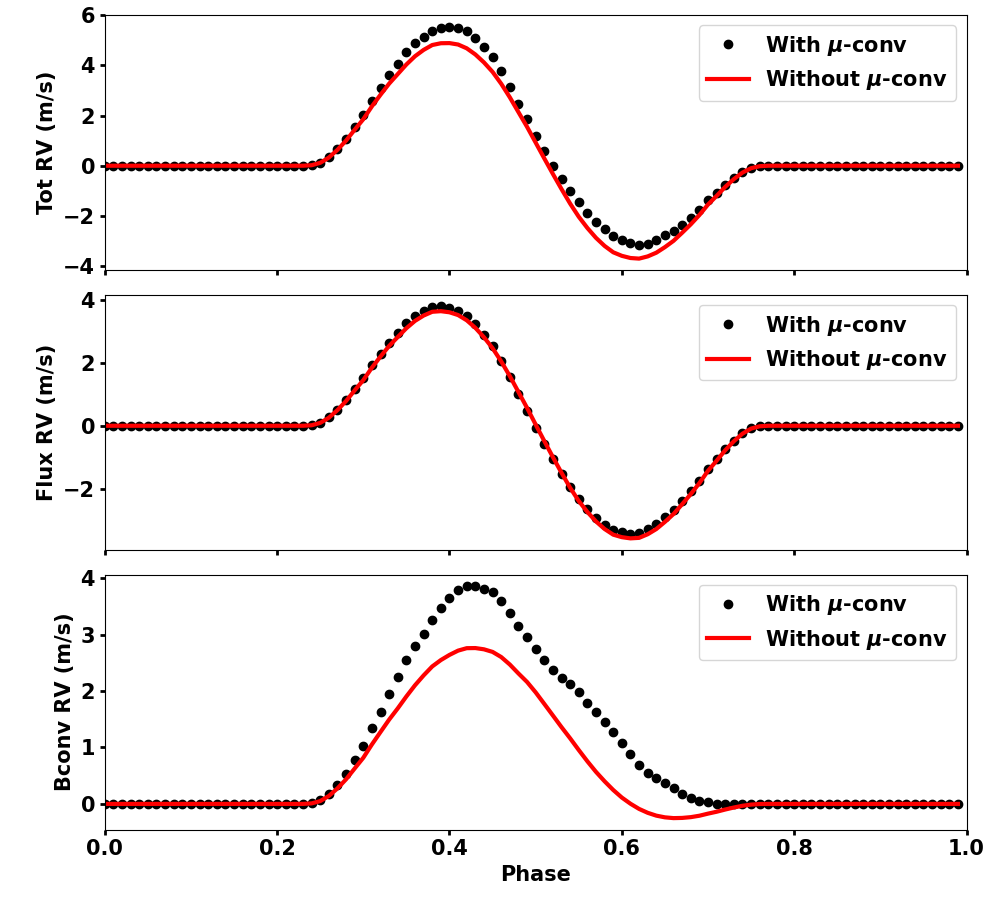}
  \caption{Comparison of the RVs derived using two different configurations for the input spectra. A single equatorial spot with 1\% area of the entire disk surface is simulated in both cases. Red line: RVs derived from simulated spectra with observed quiet sun and spot spectra without $\mu$ dependent bisector injection. Dotted line: RVs derived from simulated spectra with $\mu$ dependent bisector injection. The input spectra with $\mu$-angle dependency are generated with the Python module $Convec.model$. The RVs of the flux, CB and total effect don't change significantly when the $\mu$-dependent CB is introduced.}
    \label{Fig_BIS_solar}%
    \end{figure}

We show the RV impact of considering the $\mu$ angle dependency on the observed solar spectra in Fig.~\ref{Fig_BIS_solar}. As we can see, the impact is not significant when looking at the shape of the signal as a function of phase. This come from the fact that due to limb-darkening and the projection of active regions on the limb, most of the signal comes from larger $\mu$ angles (close to disc center). The only significant difference is for the amplitude of the CB effect. This is because we forced the CB difference between the quiet and active sun to be 340 m/s, while the CB difference between the quiet and active Kitt peak solar spectra is less than 300 m/s when measuring the average difference between the quiet and active CCF bissectors \citep[see Fig. 2 in][]{Dumusque-2014b}. We note that the complexity of modifying the bisectors depending on the center-to-limb angle is not strongly justified when using real solar spectra as input due to the small difference observed in the estimated RVs, however, this step is critical when working with synthetic spectra that does not include the proper bisectors, as described in Setc.~\ref{phoenix_spectra}.

We are conscious that depending on the magnetic field of an active region, the inhibition of the CB will be different and therefore the bisectors more or less redshifted compared to the quiet Sun, as seen in Fig.~1 in \citet{Cavallini-1985}. Also, faculae tend to have weaker magnetic fields than spots and in our case, we model those two active regions with the same bisectors and the same CB inhibition. It is therefore likely that the CB effect for faculae is slightly overestimated, and this will translate in larger RV amplitudes when modeling the CB effect for faculae. 

In summary, in this subsection we present a framework to model the Sun but also other stars (see also next subsection). Different bisectors at different $\mu$ are derived from the quiet phototsphere \citep{Lohner-Bottcher:2019aa} and facuale \citep{Cavallini-1985b} and are injected into the input spectra for which we have removed any variation in line bisector from a vertical line.

\subsubsection{Simulation based on the PHOENIX spectral database} \label{phoenix_spectra}

The implementation of convective motions described in the preceding section allow us to use synthetic spectra as input, since the effect of convection can be injected using the $Convec.model$ module. In order to study stellar activity affecting the data used in RV, a high resolution spectral library is needed. For SOAP-GPU, we decided to make it easy for the user to use as input PHOENIX high-resolution spectra \citep{Husser-2013aa}. We note however that SOAP-GPU can accept other spectral libraries, but it might be a little more difficult for the users to properly setup the inputs since the parameters to remove bisectors of input spectra are only optimized for the PHOENIX spectra and the solar atlas from the Kitt Peak Observatory FTS \citep{Wallace-2005}. The PHOENIX library propose a collection of spectra with the wavelength coverage from $500 \AA$ to $5.5 \mu m$ with resolutions of 500,000 in the optical. The library covers stellar effective temperature from 2300K to 12000K. Since the spectra in the PHOENIX library are not normalized, which is critical to perform the injection of CB described in the preceding section, we used the open source package  ``Rassine'' (\citet{Cretignier:2020ab}) to perform spectral normalization first. The continuum or spectral energy distribution (SED) of input spectra derived from ``Rassine'' are denoted as $\mathbf{SED}_{quiet}$ and $\mathbf{SED}_{active}$ respectively. 

\begin{figure}[htbp]
  \includegraphics[scale=0.35]{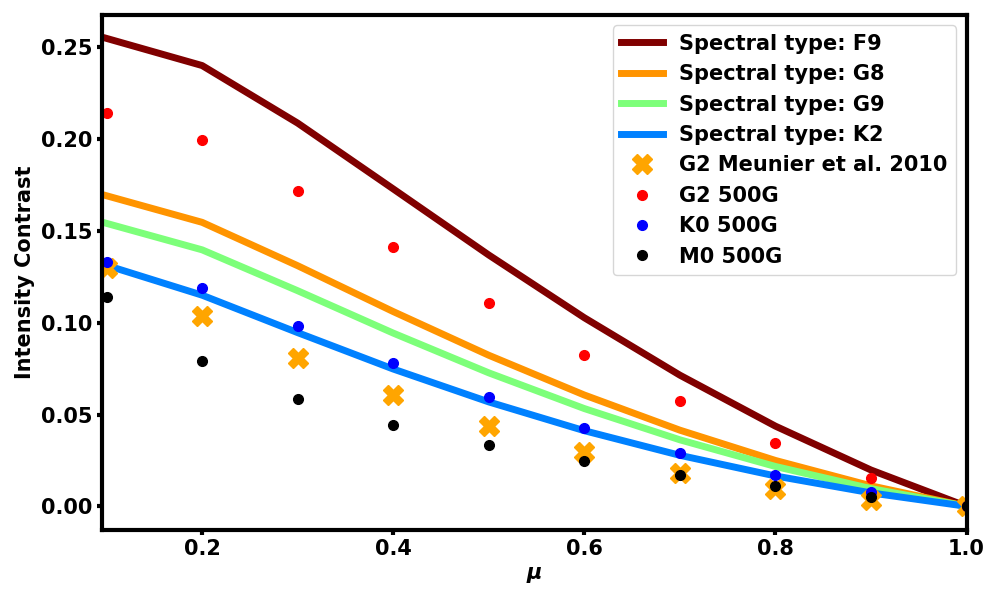}
  \caption{Faculae intensity contrast as a function of $\mu$ for different spectral types. The limb brightening curves for the G2, K0 and M0 dwarfs are derived from the parametrisation of MHD simulations for 500G faculae, shown in Table 1 in \cite{Johnson:2021mnras}. The limb brightening curves for other spectral types are linearly interpolated by using the two closet limb brightening curves. For spectral types hotter than G2, we obtain limb brightening by performing linear extrapolation using the G2 and K0 curves. The limb brightening derived from \cite{Meunier-2010a} is labeled with orange.}
    \label{limb_brightening}%
    \end{figure}

The inhibition of CB when working with PHOENIX spectra can be rewritten as:

\begin{equation}    \label{eq:bconv}   
\Delta \mathbf{S}^{'}_{bconv} (X,Y) = \mathbf{S}^{'}_{quiet, n} (X,Y) - \mathbf{S}^{'}_{active, n} (X,Y),
\end{equation}
where $\mathbf{S}^{'}_{quiet, n}$ and $\mathbf{S}^{'}_{active, n}$ are the normalized quiet and active spectra. Since the contrast between the quiet and active region is naturally included in the continuum of input PHOENIX spectra, the flux effect can be rewritten as: 
\begin{equation} \label{eq:flux}
\begin{aligned}
\Delta \mathbf{S}^{'}_{flux} (X,Y) &= \mathbf{S}^{'}_{quiet,n} (X,Y) \times \mathbf{SED}_{quiet} \\
& - \mathbf{LB} (X,Y) \times \mathbf{S}^{'}_{quiet,n} (X,Y) \times \mathbf{SED}_{active},
\end{aligned}
\end{equation}
where $\mathbf{LB}$ is the function of limb brightening. For simulation of spot regions, $\mathbf{LB} = 1.0$ along the disk. For simulation of faculae regions, SOAP 2.0 was using $\Delta T_{f} = 250.9 - 407.7\mu + 190.9\mu^2$ to model the limb brightening in temperature domain \citep{Meunier-2010a}. Since this equation is only valid for the Sun, we use the empirical equation derived from 3D MHD simulations to model other spectral types. 3D MHD simulations modelling faculae on the Sun can reproduce extremely well the limb-brightening observed for faculae \citep{Norris:2017aa}. Using similar simulations, \cite{Johnson:2021mnras} model what would be the limb-brightening on other stars (see Figure 3 and Table 1 in \cite{Johnson:2021mnras}). Using the parametrisation in \citet{Johnson:2021mnras}, we derived the limb brightening curves of faculae for a G2, K0 and M0 dwarfs. We then linearly interpolated between the G2 and K0 and K0 and M0 simulations to obtain the limb-brightening dependence for a G8 and G9 dwarf, and a K2 dwarf, respectively. To obtain the dependence for a F9 dwarf, we linearly extrapolated from the G2 and K0 models. The derived limb brightening curves from F9 to K2 are shown in Figure \ref{limb_brightening}.

Finally, the total effect from flux and convection can be derived using the following equation:
\begin{equation} \label{eq:tot}
\begin{aligned}
\Delta \mathbf{S}^{'}_{tot} (X,Y) &= \mathbf{S}^{'}_{quiet,n} (X,Y)  \times \mathbf{SED}_{quiet} \\
& - \mathbf{LB} (X,Y) \times \mathbf{S}^{'}_{active,n} (X,Y)  \times \mathbf{SED}_{active}.
\end{aligned}
\end{equation}

\begin{figure}[htbp]
  \includegraphics[scale=0.35]{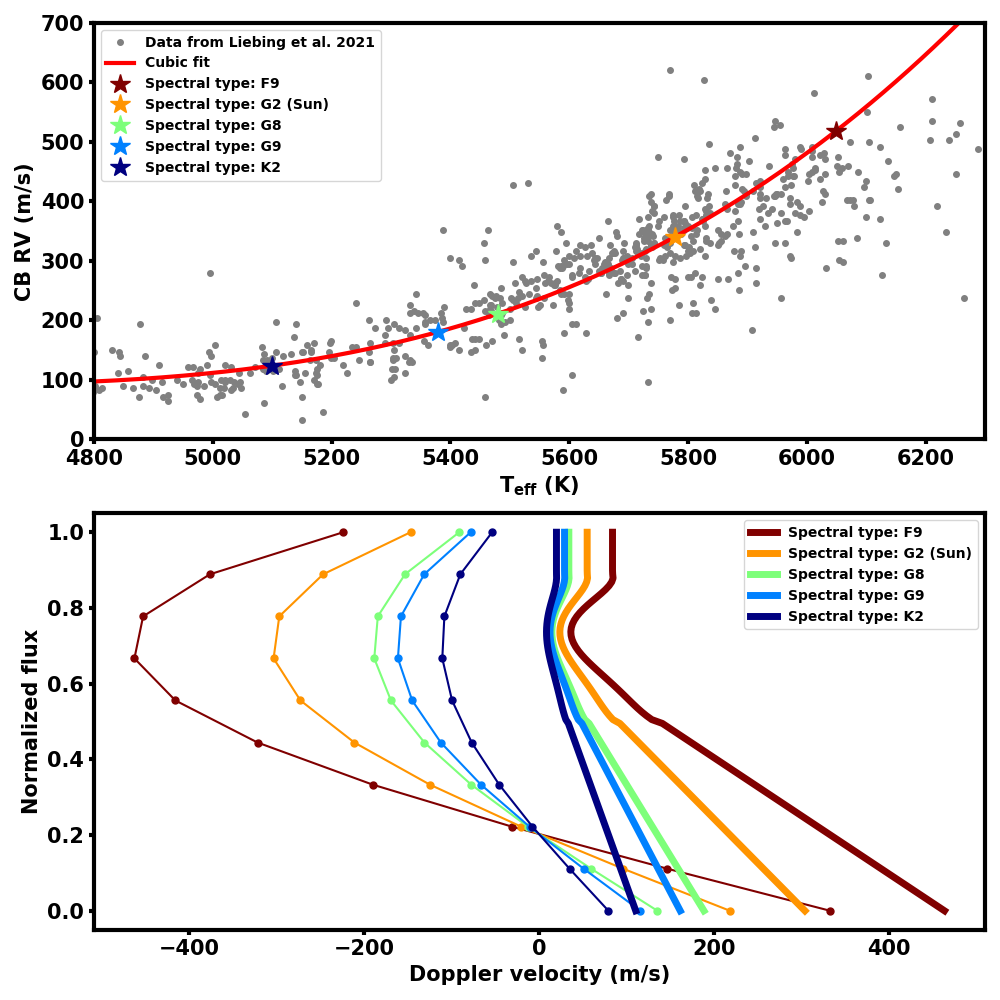}
  \caption{CB velocity as a function of spectral type. \emph{Top:} Data from \citet{Liebing:2021aa} and cubic fit to them, giving as relation  $\mathrm{CB}_{vel}=95.2388\times ((\mathrm{Teff}- 4400)/1000)^3+91.2791$. We show with coloured stars the CB velocity value for different spectral types that we model in the paper. \emph{Bottom:} The bisector of the quiet photosphere measured from \citet{Lohner-Bottcher:2019aa} at $\mu = 0.85$ (thin lines) and the bisector of an active region measured from \citet{Cavallini-1985} at $\mu = 1.0$ (thick lines) for different spectral types. We impose that the maximum difference between the quiet bisector at $\mu=0.85$ and $\mu=0.2$ (not shown here), happening at a depth of 0.58 is equal to the CB velocity given by the relation found in the top panel. We also force the difference between the quiet bisector at $\mu=0.85$ and the active bisectors, independent of their $\mu$ angle, to be equal to the same value at a depth of 0.58.}
  \label{Fig_bisector_spectral_CB}%
  \end{figure}

Convection velocity changes with spectral type, and decreases towards cooler stars than the Sun. This is a well known effect, that comes out from observations \citep[e.g.][]{Meunier:2017ab,Liebing:2021aa} and magneto hydrodynamic models of stellar phostophere \citep[e.g.][]{Allende-Prieto-2013}. As in \citet{Liebing:2021aa}, we show in the top panel of Fig.~\ref{Fig_bisector_spectral_CB} that the CB is a cubic function of effective temperature in the range 4800 to 6300 $K$. The parametrisation that we obtain is $\mathrm{CB}_{vel}=95.2388\times ((\mathrm{Teff}- 4400)/1000)^3+91.2791$. Once we have this relation to measure the velocity of CB as a function of effective temperature, we simply have to rescale the difference between the quiet Sun bisectors for $\mu=0.85$ and $\mu=0.2$ at depth 0.58 to be equal to the value given by our relation, and we also impose that the active bisectors are shifted by the same value, at the same depth (see Sect.~\ref{CB_for_mu} for justification). This allows us to model properly the change of convective velocity as a function of stellar effective temperature. We note that as in the case of the Sun, before injecting the proper bisector for the quiet and active regions, we first have to remove the bisector present in the PHOENIX spectra that we use. This is a necessary step as the PHOENIX spectral library is obtained from 1D atmospheric models and cannot properly reproduce line bisector shape. We show in the appendix, like for Fig.~\ref{Fig_CB_inhibition_kitt_peak} in the case of the Sun, how the bisector of the original PHOENIX spectra for the quiet stellar region, a faculae and a spot, are fitted before being removed. 

\begin{figure*}[htbp]
  \includegraphics[scale=0.35]{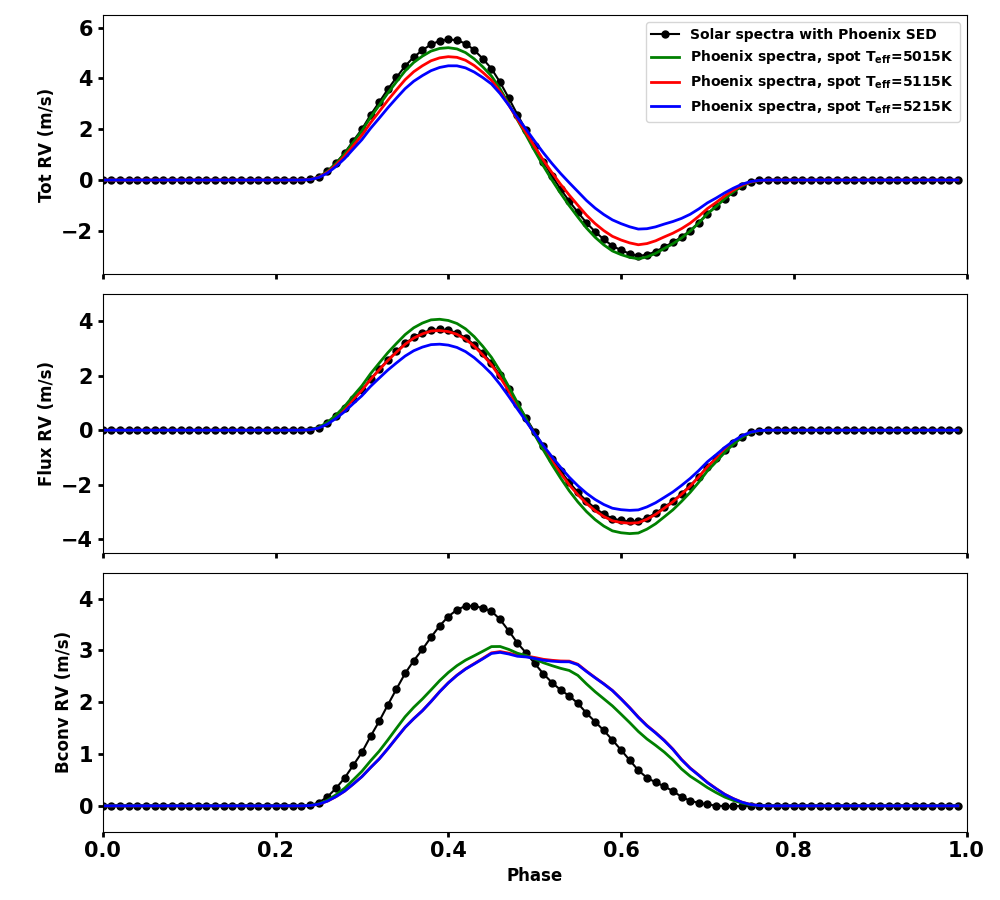}
  \includegraphics[scale=0.35]{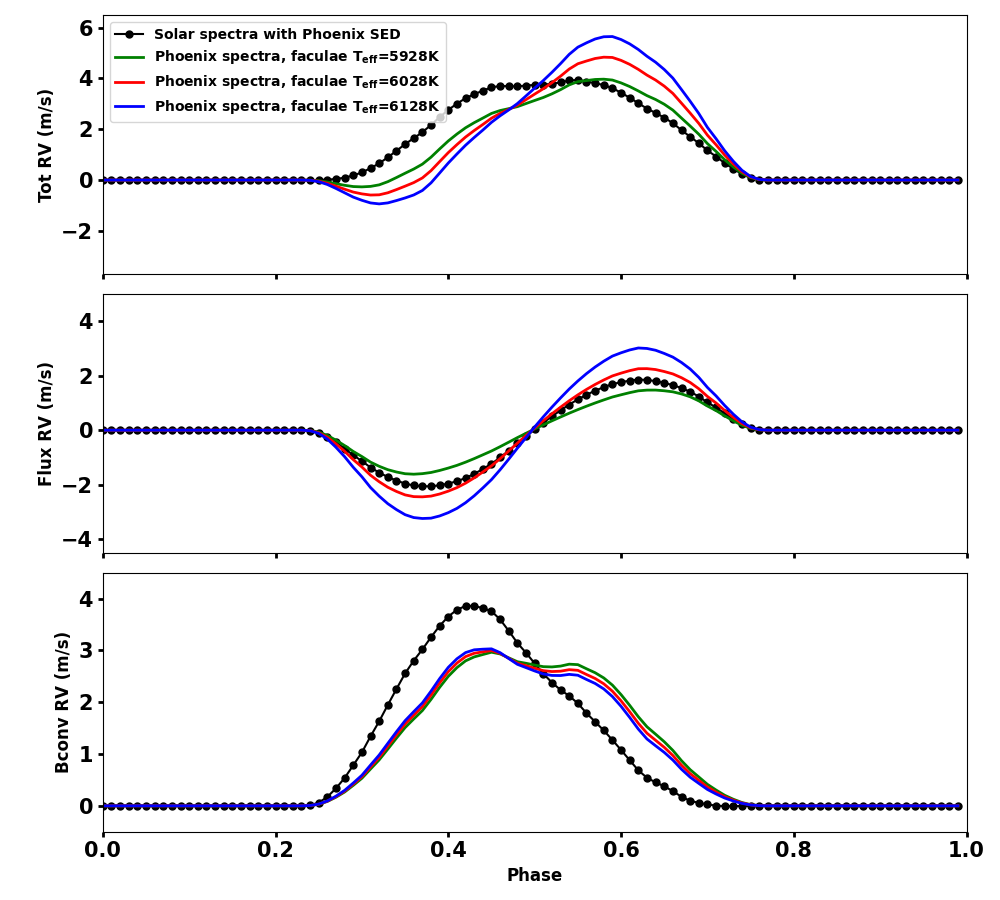}
  \caption{Simulation of an equatorial 1\% spot (left) and 1\% facula (right) with different temperatures using PHOENIX spectral library. The quiet Sun spectrum is extracted from PHOENIX spectral library with log(g) of 4.5 and $\rm{T_{eff}} = 5778 K$. The effective temperature of the spot spectra are 5015K, 5115K and 5215K, and for the facula spectra 5928K, 6028K and 6128K. We also show the results of using the observed Kitt Peak solar spectra including the PHOENIX SEDs (using Eq.~\ref{eq:flux}, \ref{eq:bconv} and \ref{eq:tot}). The $\mu$ dependent CB is included in the simulations. \emph{Left:} For the spot we see that when the difference in temperature between the spot and the quiet photosphere increases, the RV flux effect becomes larger. We note a rather large discrepancy in the CB effect between the observed solar and PHOENIX input spectra. This is likely due to the fact that the observed solar spectra are not well normalized (see Sect \ref{phoenix_spectra}). \emph{Right}: For the facula, we note the same problem of negative values for the CB effect, which expected as the same badly normalised solar active spectrum is used. We also see that considering the corresponding PHOENIX SED when using the observed solar spectra significantly change the flux contribution (see Sect \ref{phoenix_spectra}). Considering the PHOENIX SED gives results much closer to the PHOENIX simulations.}
  \label{Fig_phoenix_simu}%
  \end{figure*}

In Fig.~\ref{Fig_phoenix_simu}, we show the results of a few simulations considering $\mu$ dependent input spectra and a single 1\% equatorial spot (left panel) or a single 1\% facula (right panel). We highlight the RV outputs of SOAP-GPU when using PHOENIX spectra, with the quiet Sun temperature set to $\rm{T_{eff}} = 5778K$, the spot temperature set to $\rm{T_{eff}} = 5115K, 5015K\,\mathrm{and}\,5215K$ and the facula temperature set to $\rm{T_{eff}} = 5928K, 6028K\,\mathrm{and}\,6128K$. We also show the result when inputting the Kitt Peak solar quiet and spot or facula spectra ($\rm{T_{eff}} = 5778K$ and $5115K$ or $6028K$ at disk center, respectively) as modified in Sect.~\ref{CB_for_mu}, and including (using Eq.~\ref{eq:flux}, \ref{eq:bconv} and \ref{eq:tot}) the SED from corresponding PHOENIX spectra. As we can see, the amplitude of the RV flux effect increases with an increase in temperature difference between the quiet photosphere and the spot or facula. The CB effect does not change in amplitude with temperature difference and thus the larger amplitude observed for larger difference in temperature between quiet and active regions is solely driven by the change in contrast of the active regions with temperature. While for the flux effect, the simulation using PHOENIX spectra or observed solar spectra as input gives the same results, which is not surprising as we use the same SED, this is not the case for the CB effect. The amplitude derived show a small discrepancy of $\sim$20\%, and the maximum has a slight phase shift. This likely comes from a different flux effect contribution seen in the derived CB RV, due to molecular absorption not perfectly modeled in the PHOENIX spectra compared to solar real observations (see discussion in Sect.~\ref{chromatic_effect}). We note that this asymmetry was already something seen in the original SOAP 2.0 paper \citep[see Fig.~6 in][]{Dumusque-2014b}.
Something also interesting to note, that we see when using as input both the PHOENIX and solar spectra is the bump in the CB RV effect when the active region crosses the center of the disk (phase=0.5). This is induced by the fact that CB is maximum at $\mu=0.85$ and not at disk center, as was shown in \citet{Lohner-Bottcher:2019aa}.

\begin{figure}[htbp]
  \includegraphics[scale=0.35]{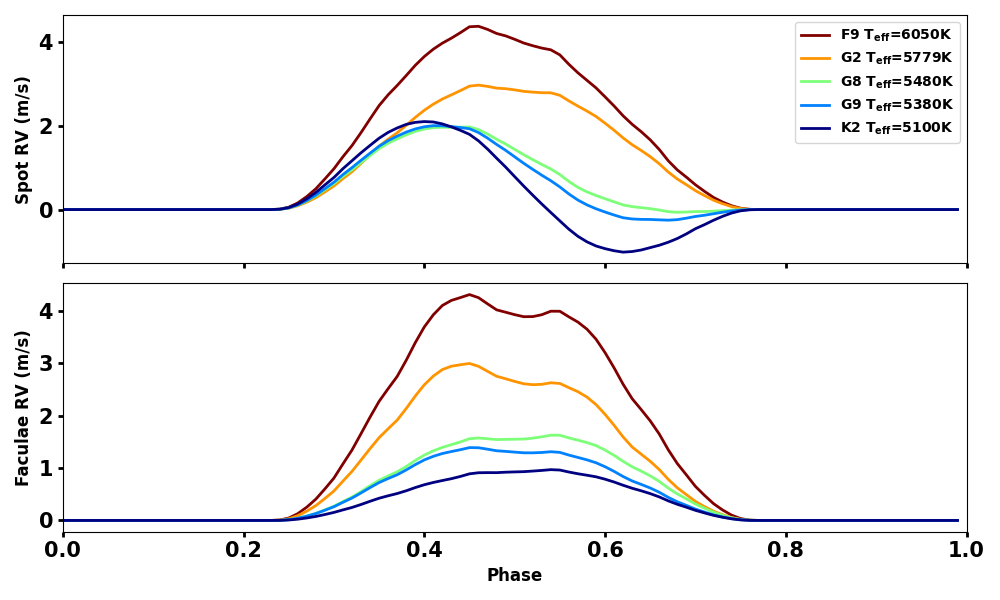}
  \caption{Simulation of the CB RV effect of an equatorial 1\% spot (top) and 1\% facula (bottom) for different temperature of the quiet photosphere (i.e. different spectral type) using the PHOENIX spectral library. The temperature difference between the quiet photosphere $\Delta \rm{T_{eff}}$ for spot and facula is fixed to 663K and 250 $K$, respectively. Due to the strong absorption effect of molecules in spots on stars cooler than the Sun, we see the appearance of a flux effect in the derivation for the CB effect only (see Sect \ref{phoenix_spectra}).}
  \label{Fig_phoenix_CB}%
  \end{figure}

In Fig.\ref{Fig_phoenix_CB} we show the result of the estimated CB RV effect for an equatorial 1\% spot or facula for stars of different temperature (i.e. spectral type): 6050 $K$ (F9), 5778 $K$ (G2), 5480 $K$ (G8), 5380 $K$ (G9) and 5100 $K$ (K2). For the spot, we see a positive only effect for the F9 and G2 simulations, which is expected, but for later spectral type, we start to see the emergence of a flux effect. This effect, as already discussed in Sect.\ref{chromatic_effect} comes from the absorption of molecules, that change significantly over a few hundreds of Kelvin for the quiet photosphere at 5480 $K$ and a spot at 4817 $K$ for the G8 simulation for example. Also, we see that the more we go towards cooler stars, the more the CB effect show a flux contribution, and this comes from the fact that molecular absorption is not linear with effective temperature. Although molecular absorption prevent us of clearly separating the flux effect of spots from the CB effect like in the case of the F9 or G2 star, the total RV effect including both contributions is still properly estimated. Therefore, users should be careful when interpreting the estimated RV induced by the inhibition of convection for spots on stars with spectral type later than the Sun, however, they can trust the total RV effect estimated.
Regarding the facula, we observe a positive only effect for all spectral type, and thus the CB effect is properly modeled for this type of active regions.

\subsubsection{Limitation of SOAP-GPU when moving away from solar twins} \label{SOAP_limitations}

On a careful note, we want to warn the user that a lot of the physics included in SOAP-GPU is based on solar observations, like for example the variation of the bisector of the quiet photosphere with respect to the center-to-limb angle \citep[][]{Lohner-Bottcher:2019aa}, or the bisector of an active region extracted from solar observations \citep[][]{Cavallini-1985}. Therefore, although we try to correct for some effects, like the variation of CB velocity as a function of spectral type, the more we go away from the solar case, the more we should be careful about interpreting the results coming out of SOAP-GPU. 

We note that when modifying the bisector shape of solar or PHEONIX spectra for the disk center, to account for different convection velocity across spectral type, we model the effect of the “third signature” of granulation \citep[][]{Gray-2009}. The inherent shape of the bisector (known as the “second signature” of granulation) and how it varies with $\mu$ \citep[][]{Lohner-Bottcher:2019aa} is still the one observed for the Sun and it is well known in the literature that the bisector shape of disc-integrated observations, and therefore by analogy at disk center, varies significantly among luminosity class and spectral type \citep[see Figure 17.15 in][]{Gray-2008, Basturk:2011aa}. In those papers, we see that inside the small range from early G’s dwarfs to early K’s dwarfs, the bisector shapes does not change drastically and therefore, although the integrated spectra for those stars won't be realist in terms of line bisector, the way SOAP-GPU models stellar activity should still be quite realistic as in the end, what counts is the differential between the activity and quiet phases. As we can see in the preceding subsection, besides SOAP-GPU not being able anymore to separate the inhibition of the CB effect from the flux effect for early K's, the tests we performed show that the code seems to behaves quite well in estimating the total RV effect induced by spots and faculae. However, outside of this small range from G to K dwarfs, bisectors are completely different, and we warn the users that the present version of SOAP-GPU might give very unrealistic stellar integrated spectra and estimation of stellar activity. There is perhaps only one exception for dwarfs cooler than early K’s, for which the velocity of convection decreases to level that are very difficult to measure. For those stars, stellar activity is dominated by the flux effect from spot and faculae. Therefore, for such stars, users should ignore the output from the CB effect and only consider the flux effect.

To better model stellar activity for stars different than the Sun, we would require disk-resolved spectra for other stars. Those could come from 3D MHD simulations \citep[e.g.][]{Johnson:2021aa, Dravins:2021aa}, or thanks to spatially resolved spectroscopic observation across stellar surfaces thanks to transiting planets \citep[e.g.][]{Dravins:2017aa}. Both approach are challenging, however could led to a much more realistic modelisation of stellar activity from other stars than our Sun. We note that it is rather easy to input different spectra in SOAP-GPU than the ones provided (the Sun and PHEONIX spectra), and therefore if users have access to disk-resolved spectra with more realistic line-bisectors, SOAP-GPU could still be used to obtain efficiently disk-integrated spectra and model the corresponding stellar activity effect. We note that the CB injection part is an individual module in SOAP-GPU that users can easily modify.

\section{Conclusions}\label{sec5}

In this paper, we present a GPU-based improvement to SOAP 2.0, named SOAP-GPU, that allows to efficiently model stellar activity at the spectral level. With the implementation of GPU interpolation and summation, benchmark calculations demonstrate that SOAP-GPU improves the computational speed by a factor of 60 when modeling stellar activity on a full visible spectral range at R=115’000 of resolution, while having the same accuracy.

Beside the huge gain in speed, SOAP-GPU also provides a more complex modelisation of stellar activity compared to SOAP 2.0. Complex active region scenarios, with regions overlapping is now handled by the code. This is mainly useful when modeling active phases of a star like the Sun, with hundreds of active regions. The contrast of the active regions is now wavelength dependant, and therefore change for each wavelength of the modeled spectra. The dependence of line bisector with center-to-limb angle $\mu$, following the work of \citet{Lohner-Bottcher:2019aa} and \citet{Cavallini-1985}, is also now accounted for for the Kitt Peak observed quiet and active atlases, but also for PHOENIX spectra. Although the induced RV effect is rather negligible when using the observed solar spectra as input, including the framework to change line bisector is crucial to properly model convection when injecting PHOENIX spectra as input. The use of PHOENIX spectra allows us now to model a wide variety of stars with different stellar and active region properties, and allows us as well to better model faculae, as the corresponding spectrum now has the proper effective temperature (SOAP 2.0 was using the Kitt Peak sunspot atlas to model faculae).

When modeling the inhibition of CB effect using as input the solar Kitt Peak quiet and active spectral atlases, we noticed that the derived RVs go negative, which is not expected. This comes from molecular absorption that can be seen in the spot spectrum due to lower temperature compared to the quiet Sun. Even though we do not include the contrast of the active region when modeling the RV CB effect only (see Eqs.~\ref{eq:2} and ~\ref{eq:bconv}), the difference in flux at the level of molecular absorption bands will show up as a flux effect in the estimated CB RVs. Positive values will be added to the CB RV effect before the spot crosses the stellar center, and negative values after, therefore creating an asymmetry.

When modeling other stars than the Sun using PHOENIX spectral library, users should be aware that a lot of physics included in SOAP-GPU are based on solar observations, and although the code tries to correct for known effects like the variation of CB velocity as a function of effective temperature (i.e spectral type, see Sect.~\ref{phoenix_spectra}), the more we go away from the Sun, the more the results should be interpreted with caution (see discussion in Sect.~\ref{SOAP_limitations}). The modelisation of stellar activity for other stars than the Sun is currently limited by the knowledge we have about disk-resolved bisectors for such stars. Such information is very challenging to obtain, however, 3D MHD simulations \citep[e.g.][]{Johnson:2021aa, Dravins:2021aa} and resolved spectroscopic observation of other stars due to planetary transits \citep[e.g.][]{Dravins:2017aa} could significantly help.

Also, when modeling stars of later spectral type than the Sun, we are not able anymore to separate clearly the inhibition of CB effect from the flux effect due to the strong absorption of molecules. However, the output for the total RV effect (flux plus inhibition of CB effects) should be modeled properly. SOAP-GPU have been tested up to a K2 star (T$_\mathrm{eff} = 5100\,K$) with spots 663 $K$ cooler and give satisfactory results. Modeling later spectral type is challenging mainly due to continuum normalisation of the PHOENIX spectra and injection of spectral line bisector due to line blending, and users should be very careful about the interpretation of the results for such stars with the present code.

There are still some improvements that could be made to better model the physics at play. Although the spot and facula spectrum used when considering PHOENIX spectra as input are of different temperature, and therefore in spectral content, we still associate to those regions the same active bisector as measured for the Sun on a facula \citep[][]{Cavallini-1985}. Spots are induced by stronger magnetic fields than facula, and thus it is likely that the bisector of spectral lines will be slightly different. Although it is possible to know what is the bisector of a few spectral lines inside a spot at disk center, to our knowledge, no measurement of spot line bisectors for different $\mu$ angles are published. \citet{Cavallini-1985} also show in their Fig.~1 that depending on the facula observed, the bisector shape changes due likely to different magnetic field strength and therefore different level of CB inhibition. As can be seen in Fig.~\ref{Fig_BIS_solar}, the effect of inducing $\mu$ dependant spectral line shape in the quiet and active regions is rather small, and although with more solar data about spots and faculae we could better model the physics at play, results in terms of RV derivation would be rather similar. This likely comes from the fact due to limb-darkening, most of the weight is put on the disk center, where spectral lines does not change significantly in shape.

With the performance of SOAP-GPU, it is now possible to model activity at the spectral level for complex stellar surfaces with many active regions and for a long period of time. A solar activity simulator, either based on statistical properties of solar active regions \citep[similar to][]{Borgniet-2015aa} or on the observed distribution of those \citep[similar to e.g.][]{Meunier-2010a} will be published in a forthcoming paper. We encourage any person working on techniques to separate the activity effect from planetary signals at the spectral level, to test their framework on SOAP-GPU simulations, where photon-noise, instrumental and telluric systematics are not perturbing the spectral timeseries.

\begin{acknowledgements}
We thank the anonymous referee for the insightful and constructive comments on this paper. We thank Michael Crerignier for his help in normalizing PHOENIX spectra with RASSINE. We also thank Xiang Gao for the constructive comments on GPU computing.
This project has received funding from the European Research Council (ERC) under the European Union’s Horizon 2020 research and innovation programme (grant agreement SCORE No 851555).
This work has been carried out within the framework of the National Centre of Competence in Research PlanetS supported by the Swiss National Science Foundation. The authors acknowledge the financial support of the SNSF.
\end{acknowledgements}

\bibliographystyle{aa}
\bibliography{SOAP_GPU}

\begin{appendix}

\section{Line bisectors of PHOENIX spectra}\label{sec_App}

\begin{figure*}[htbp]
  \centering
  \includegraphics[scale=0.45]{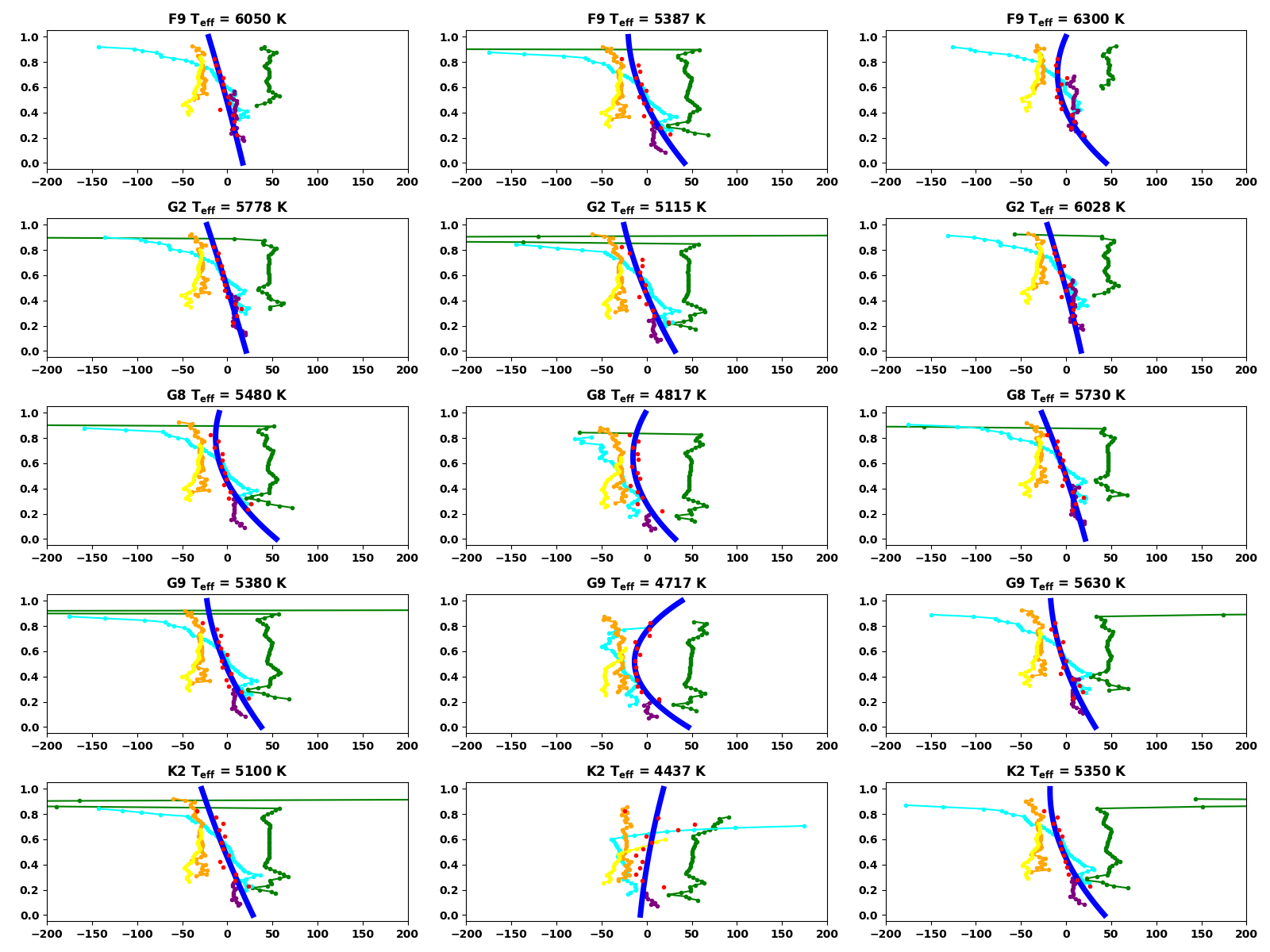}
  \caption{Bisector of PHOENIX spectra. For each input seed spectrum using PHOENIX spectral library, we use five strong iron lines: FeI $5250.2084\AA$ (green), FeI $5250.6453 \AA$ (cyan), FeI $5434.5232\AA$ (purple), FeI $6173.3344\AA$ (orange) and FeI $6301.5008\AA$ (yellow) to measure the average bisector of the input spectra. Bisector outliers outside a window of $0.1 \AA$ around each line center are rejected to avoid those points, certainly affected by line blending, to bias our measurement of line bisector. Each line correspond to a different spectral type, and from left to right, we can see the bisector of the spectrum used for the quiet photosphere, a spot region (663 $K$ cooler) and a facula region (250 $K$ hotter). We average those line bisectors at certain depth (as shown by the red dots) and fit the obtained data with a second order polynomial. The fitted bisector is used to remove the bisector of input seed spectrum.}
  \label{Fig_phoenix_bisfit}%
  \end{figure*}

As discussed in Sects.~\ref{CB_for_mu} and \ref{phoenix_spectra}, before injecting the $\mu$ dependant bisector for solar or PHOENIX spectra to properly model CB and its inhibition close to the limb and in active regions, we need to remove any bisector shape already present in the input spectra. As PHOENIX spectral library is generated from 1D spectral synthesis, the line bisectors cannot include properly the CB effect and therefore should be close to straight. In Fig.~\ref{Fig_phoenix_bisfit}, we show for each simulation of different spectral types the bisector of a few iron lines that are used in \citet{Lohner-Bottcher:2019aa}. For each spectral type simulated, we show the bisectors for the quiet photosphere, but also for simulated spot and faculae, 663 $K$ cooler or 250 $K$ hotter, respectively. As expected, most of the bisector are close to straight lines. We however fitted the average bisector with a second order polynomial to remove the small curvatures observed before injecting the proper bisectors at different $\mu$ angles (see Sect.~\ref{CB_for_mu}). It is not clear if those curvatures are real effect in the spectral synthesis, or simply due to blends. The correction performed is small compared to the bisectors that we inject afterward, therefore if only due to blends, this process does not significantly change the outputs.

\end{appendix}

\end{document}